\documentclass{article}%
\usepackage{amssymb}
\usepackage{amsfonts}
\usepackage{amsmath}
\usepackage{algorithmic}
\usepackage{float}
\usepackage{graphicx}%
\providecommand{\U}[1]{\protect\rule{.1in}{.1in}}
\newtheorem{theorem}{Theorem}

\newtheorem{algorithm}[theorem]{Algorithm}

\newtheorem{proposition}[theorem]{Proposition}

\begin{document}

\date{}
\title{Epidemic threshold in directed networks }
\author{Cong Li, Huijuan Wang and Piet Van Mieghem\\Faculty of Electrical Engineering, Mathematics and Computer Science, \\Delft University of Technology, Delft, The Netherlands}
\maketitle

\begin{abstract}
Epidemics have so far been mostly studied in undirected networks. However,
many real-world networks, such as the online social network Twitter and the
world-wide web, on which information, emotion or malware spreads, are directed
networks, composed of both unidirectional links and bidirectional links. We
define the \emph{directionality }$\xi$ as the percentage of unidirectional
links. The epidemic threshold $\tau_{c}$ for the
susceptible-infected-susceptible (SIS) epidemic is lower bounded by
$1/\lambda_{1}$\ in directed networks, where $\lambda_{1}$, also called the
spectral radius, is the largest eigenvalue of the adjacency matrix. In this
work, we propose two algorithms to generate directed networks with a
given\textbf{ }directionality $\xi$. The effect of $\xi$ on the spectral
radius $\lambda_{1}$, principal eigenvector $x_{1}$, spectral gap
$(\lambda_{1}-\left\vert \lambda_{2}\right\vert )$ and algebraic connectivity
$\mu_{N-1}$ is studied. Important findings are that the spectral radius
$\lambda_{1}$ decreases with the directionality $\xi$, whereas the spectral
gap and the algebraic connectivity increase with the directionality $\xi
$.\ The extent of the decrease of the spectral radius depends on both the
degree distribution and the degree-degree correlation $\rho_{D}$.
Hence,\textbf{\ }in directed networks, the epidemic threshold is larger and a
random walk converges to its steady-state faster than that in undirected
networks with\ the same degree distribution.

\end{abstract}

\section{Introduction}

Much effort has been devoted to understand epidemics on networks, mainly
because of the increasing threats from cybercrime and the expected outbreak of
new fatal viruses in populations. Epidemics have been studied on undirected
networks for a long time and many authors (see \cite{Bailey, Barrat,
Castellano, Daley, Kephart, Pastor-Satorras, Van Mieghem_Virus}) addressed the
existence of an epidemic threshold $\tau_{c}$ in the
susceptible-infected-susceptible\textbf{\ }(SIS) epidemic process\textbf{\ }%
\cite{Anderson}. We consider the SIS epidemic process in an undirected network
$G(N,L)$, characterized by a symmetric adjacency matrix $A$ consisting of
elements $a_{ij}$\ that are either one or zero depending on whether node
$i$\ is connected to $j$\ or not. Each node $i$ has two possible states at
time $t$: either $X_{i}(t)=0$, meaning that the node is healthy and
susceptible, or $X_{i}(t)=1$, for an infected node. Initially, a certain
percentage of nodes are randomly selected as infected. The infection from an
infected node to each of its healthy neighbors and the curing of an infected
node are assumed to be independent Poisson processes with rates $\beta$ and
$\delta$, respectively. Every node $i$ at time $t$ is either infected, with
probability $v_{i}(t)=$\ \textrm{Prob}$[X_{i}(t)=1]$\ or healthy (but
susceptible) with probability $1-v_{i}(t)$. This is the general
continuous-time description of the simplest type of a SIS epidemic process on
a network. The epidemic threshold $\tau_{c}$ separates two different phases of
the epidemic process on a network: if the effective infection rate
$\tau\triangleq\beta/\delta$\ is above the threshold, the infection spreads
and becomes persistent; if $\tau<\tau_{c}$, the infection dies out
exponentially fast. The epidemic threshold $\tau_{c}^{(1)}=\frac{1}%
{\lambda_{1}}$ of the N-intertwined Mean-field Approximation (NIMFA) \cite{Van
Mieghem_Virus, Piet3, Piet_SIRandSIS, Wang} of the SIS model lower bounds the
real threshold, where $\lambda_{1}$ is the largest eigenvalue of the adjacency
matrix $A$, also called the spectral radius.

Topologies of undirected networks have been mostly modeled by Erd\H{o}s and
R\'{e}nyi\footnote{An Erd\H{o}s-R\'{e}nyi random graph can be generated from a
set of $N$ nodes by randomly assigning a link with probability $p$ to each
pair of nodes.} \cite{Erdos_Renyi1959, Gilbert, Erdos1960} as binomial
networks, by B\'{a}rabasi and Albert\footnote{A B\'{a}rabasi-Albert graph
starts with $m$ nodes. At every time step, we add a new node with $m$ links
that connect the new node to $m$ different nodes already present in the graph.
The probability that a new node will be connected to node $i$ in step $t$ is
proportional to\ the degree $d_{i}(t)$ of that node. This is referred to as
preferential attachment.} \cite{Barabasi1999} as power-law networks, or by
Watts and Strogatz\footnote{A Watts-Strogatz small-world graph can be
generated from a ring lattice with $N$ nodes and $k$ edges per node, by
rewiring each link at random with probability $p$.} \cite{Watts1998}
as\textbf{\ }small-world networks. More complicated static and dynamic models,
such as the configuration model \cite{Bollobas, Molloy, Newman2001}, are also
proposed to approximate real-world networks. However, many real-world networks
are \emph{directed} networks, some examples are shown in Table
\ref{realworld_table1}. The dataset of the real-world networks is obtained
from \cite{Boldi2004, Boldi} and the description of these networks is attached
in the Appendix A.%

%TCIMACRO{\TeXButton{beg_table}{\begin{table}\centering\caption
%{Percentage of unidirectional links in real-world networks }\label
%{realworld_table1}}}%
%BeginExpansion
\begin{table}\centering\caption
{Percentage of unidirectional links in real-world networks }\label
{realworld_table1}%
%EndExpansion

\bigskip

$%
\begin{array}
[c]{cccc}\hline\hline
\text{Real-world networks} & N & L_{arcs} & \xi\\\hline
\text{Enron} & 69\,,244 & 276\,,143 & 84.29\%\\
\text{Ljournal-2008} & 5\,,363,\,260 & 79\,,023\,,142 & 25.32\%\\
\text{Twitter-2010} & 41,\,652,\,230 & 1\,,468\,,365,\,182 & 64.29\%\\
\text{WordAssociation-2011} & 10\,,617 & 72,\,172 & 76.77\%\\
\text{cnr-2000} & 325,557 & 3,216,152 & 70.33\%\\
\text{in-2004} & 1,382,908 & 16,917,053 & 60.68\%\\
\text{eu-2005} & 862,664 & 1,935,140 & 67.80\%\\
\text{uk-2007-05@100000} & 100,000 & 3,050,615 & 82.23\%\\
\text{uk-2007-05@1000000} & 1,000,000 & 41,247,159 & 79.71\%\\\hline\hline
\end{array}
$%
%TCIMACRO{\TeXButton{beg_table}{\end{table}}}%
%BeginExpansion
\end{table}%
%EndExpansion

Two kinds of links, namely bidirectional links and unidirectional links, exist
in directed networks. If node $i$ is connected to node $j$ (denoted by
$i\rightarrow j$) then $j$ is also linked to $i$ (denoted by $j\rightarrow
i$), one bidirectional link exists between nodes $i$ and $j$; and if either
$i\rightarrow j$ or $j\rightarrow i$ exists, but not both in between the node
pair $i$ and $j$, a unidirectional link exists. Here, we define the
directionality as $\xi=L_{unidirectional}/L_{arcs}$, where the number of arcs
(the number of $1$-elemets in the adjacency matrix) $L_{arcs}=\sum_{i}\sum
_{j}a_{ij}=u^{T}Au$, ($u$ is the all-one vector), can also be calculated by
$L_{arcs}=L_{unidirectional}+2L_{bidirectional}$. A directed network with
directionality $\xi$ is denoted by $G^{(\xi)}$. The network $G^{(\xi=0)}$ is a
bidirectional network or an undirected network, whose adjacency matrix is
symmetric, when $\xi=0$. The network $G^{(\xi=1)}$ is a directed network
without any bidirectional link, when $\xi=1$.\textbf{\ }A high directionality
is observed in Twitter, as shown in Table\textbf{\ }\ref{realworld_table1}. A
link runs from user A to user B if user A follows user B in Twitter, where
user A is called the "follower" of user B. The fact that user A "follows" user
B, does not necessarily mean that the reverse is also true. For example, a
famous person could have millions of followers but he/she may not follow many
others. This explains the high directionality $\xi$ of Twitter. The
virtual-community social networks, such as LiveJournal, have a low
directionality (see Table \ref{realworld_table1}), mainly because they aim to
construct virtual connections in between real-life friends, and friendship
relations are usually mutual.

There has been an increasing interest in the study of directed networks.
Topological properties of directed networks, such as the short loops, closure
connectivity, degree, domination and communities on realistic directed
networks have already been studied in \cite{Bianconi, Robins, Kim, Brink,
Albert, Newman_SIAM}. Garlaschelli and Loffredo \cite{Garlaschelli}%
\ investigated the reciprocity \cite{Newman_book} in directed networks, where
the reciprocity is equal to $1-\xi$. Processes taking place on networks, such
as synchronization, percolation and epidemic spread, have also been researched
\cite{Garlaschelli, Park, Schwartz, Dorogovtsev, Newman_virus in email} in
real directed networks. Percolation theory for directed networks\ with $\xi
=1$\ was\ firstly developed by Newman et al. \cite{Newman2001, Callaway}.
Then, Bogu\~{n}\'{a} and Serrano \cite{Boguna} pointed out that even a small
fraction of bidirectional links suffices to percolate the network. Moreover,
Meyers et al. \cite{Meyers} used a generating function method to predict
the\ epidemic threshold in directed networks with $\xi<1$\ and the size of the
infected cluster. Recently, Van Mieghem and van de Bovenkamp have proven that
the NIMFA epidemic threshold $\tau_{c}^{(1)}=\frac{1}{\lambda_{1}}$ of the SIS
epidemic process also holds for directed networks \cite{Piet_SIRandSIS}.
Stimulated by the directed social networks with different directionalities,
here, we focus on the influence of the directionality $\xi$ on the epidemic
threshold $\tau_{c}^{(1)}=\frac{1}{\lambda_{1}}$ and other spectral properties.

This paper is organized as follows. In Section \ref{Algorithms}, we propose
two algorithms that could be applied to a bidirectional network to generate a
directed network with an arbitrary given directionality $\xi$, by rewiring or
resetting links. The in- and out- degree distribution of the generated
directed network is the same as the degree distribution of the original
bidirectional network. Chen and Olvera-Cravioto \cite{Chen} proposed an
algorithm to generate a directed network with a given in- and out-degree
distribution, which is similar to the configuration model. However, their
algorithm in \cite{Chen} cannot generate a directed network with a given
directionality $\xi$. In Section \ref{Sec_spectral}, we investigate the effect
of the directionality $\xi$ on the spectral radius $\lambda_{1}$, the
principal eigenvector $x_{1}$ (the eigenvector corresponding to $\lambda_{1}%
$), the spectral gap $\lambda_{1}-|\lambda_{2}|$ and the algebraic
connectivity $\mu_{N-1}$ in both directed binomial\footnote{For example, an
Erd\H{o}s-R\'{e}nyi random network is a binomial network with the Pearson
degree correlation $\rho_{D}=0$. A general binomial network could possibly
have an assortativity $\rho_{D}$ within a large range.} and\ power-law
networks, whose in-degree and out-degree both follow a binomial (or power-law)
distribution.\ Interestingly, we find that the spectral radius $\lambda_{1}$
of networks $G^{(\xi=0)}$ is larger than that of directed networks
$G^{(\xi=1)}$ when the degree distribution and the assortativity of these
networks are the same. This means that the epidemic\ threshold $\tau_{c}$ in
undirected networks is smaller than that in directed networks with the same
degree distribution and assortativity. Furthermore, we explore the influence
of the Pearson degree correlation coefficient $\rho_{D}$ (also called the
assortativity) on the epidemic threshold $\tau_{c}^{(1)}$ in both directed
binomial and directed power-law networks with different $\xi$, in Sec.
\ref{Section_infect_assortativity}. The $\rho_{D}$ is the Pearson correlation
coefficient of degrees \cite{Newman_Ass} at either ends of a link and lies in
the range $[-1,1]$. Actually, there are four degree correlations, namely the
in-degree and in-degree correlation, the in-degree and out-degree correlation,
the out-degree and in-degree correlation and the out-degree and out-degree
correlation, in directed networks. We consider directed networks where the
in-degree and out-degree of each node are the same. In this case, the four
degree correlations are equal to each other and can be all referred as the
degree correlation (or the assortativity). The decrease of the spectral radius
$\lambda_{1}$ with $\xi$ is large when the assortativity $\rho_{D}$ is large
in directed binomial networks, whereas the opposite is observed in directed
power-law networks.

\section{Algorithm Description \label{Algorithms}}

The networks mentioned in this paper are simple, without self-loops and
multiple links from any node to any other node. Here,\ we propose two
algorithms, In-degree and Out-degree Preserving Rewiring Algorithm (IOPRA) and
Link resetting algorithm (LRA), which both can be applied to any network to
generate a directed network with a given directionality $\xi$.\ In this study,
we only apply these two algorithms to generate directed networks with the same
in- and out- degree distribution. The difference is that IOPRA preserves the
in- and out- degree of each node, while, LRA may change the in- and out-degree
of any node. IOPRA is inspired by the degree preserving rewiring, which has
been presented in \cite{Van Mieghem, Maslov}. We firstly introduce the
degree-preserving rewiring, which monotonously increases or decreases the
assortativity $\rho_{D}$, while maintaining the node degrees unchanged, in
undirected networks. Afterwards, we describe IOPRA and LRA in detail.

\subsection{Degree-preserving rewiring}

The degree-preserving rewiring \cite{Van Mieghem} can either increase or
decrease the assortativity of a bidirectional network: (a) the
degree-preserving assortative rewiring: randomly select two links associated
with four nodes and then rewire the two links such that the two nodes with the
highest degree and the two lowest-degree nodes are connected\textbf{,
}respectively. If any of the newly rewired links exists before rewiring,
discard this step and a new pair of links is randomly selected; (b) the
degree-preserving disassortative rewiring: randomly select two links
associated with four nodes and then rewire the two links such that the
highest-degree node and the lowest-degree node are connected,\ and the
remaining two nodes are also connected, as long as the newly rewired links do
not exist before rewiring. Either rewiring step (a) or (b) can be repeated to
monotonically increase or decrease the assortativity in a bidirectional network.

\subsection{In-degree and out-degree preserving rewiring algorithm (IOPRA)}

\begin{figure}[h]%
\centering
\includegraphics[
height=1.7262in,
width=3.9704in
]%
{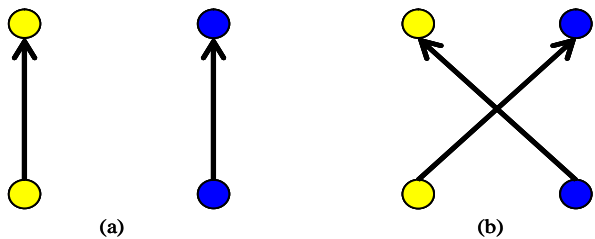}%
\caption{(Color online) In-degree and out-degree preserving rewiring}%
\label{picture1}%
\end{figure}

IOPRA can be applied to change the directionality of networks. We define our
In-degree and Out-degree Preserving Rewiring Algorithm (IOPRA) as follows:
randomly choose two unidirectional links with four end nodes, and rewire the
two unidirectional links. In IOPRA, the head of one unidirectional link only
can rewire with the head of the other unidirectional\ link, in order to
maintain both the in-degree and out-degree of the four nodes unchanged (see
Figure \ref{picture1}). We don't rewire if such rewiring can introduce
duplicated links from any node to any other. We discard the rewiring step if
this rewiring step doesn't change the directionality $\xi$ towards the given
directionality. In both cases, we randomly reelect a pair of unidirectional
links associated four nodes. We illustrate the process of IOPRA changing the
directionality in Algorithm \ref{IOPRAAlgo} (see Appendix C). IOPRA actually
changes the directionality $\xi$ of a given network $G$ without changing the
in- and out-degree of each node. If the original network $G$ is an undirected
network, the in-degree sequence is exactly the same as the out-degree sequence
in the directed network $G^{(\xi)}$ generated by IOPRA.
%TCIMACRO{\FRAME{fhFU}{3.9704in}{1.7262in}{0pt}{\Qcb{(Color online) In-degree
%and out-degree preserving rewiring}}{\Qlb{picture1}}{picture1.eps}%
%{\special{ language "Scientific Word";  type "GRAPHIC";  display "USEDEF";
%valid_file "F";  width 3.9704in;  height 1.7262in;  depth 0pt;
%original-width 6.5112in;  original-height 3.141in;  cropleft "0";
%filename 'Picture1.eps';file-properties "XNPEU";}}}%
%BeginExpansion
%EndExpansion

IOPRA changes the directionality, as well as randomizing the connections of
the original network, without changing the degree of any node. Hence, if the
initial network\ is a random network, e.g. an Erd\H{o}s-R\'{e}nyi (ER) network
or a scale-free network generated by the configuration model, where the
connections are originally laid at random, IOPRA changes only the
directionality $\xi$. However, if we apply IOPRA to\ a nonrandom network, e.g.
a lattice, the resulting network has not only a different directionality but
also a more randomized structure.

\subsection{Link resetting algorithm (LRA)\label{LRAalgorithm}}

We start with\ a bidirectional network $G$, and use the Link Resetting
Algorithm (LRA) to change the directionality (see Algorithm\ \ref{LRAAlgo} in
Appendix C). We randomly choose a fraction $\xi$ of the bidirectional link
pairs from $G$. Then, we randomly choose only one unidirectional link from
each bidirectional link, and randomly relocate the selected unidirectional
links to a place without any link.\ In this work, we only apply LRA to ER
networks. In this case, the in- and out-degree of the generated directed
network follow the same binomial degree distribution\ as the original network.
However, the in-degree and out-degree of any node in the generated
network\ may differ from those in the original network $G$. When LRA is
applied to other types of networks, such as the power-law networks, the
original in- and out-degree distributions are destroyed, and tend to be binomial.

In summary, two types of directed binomial networks can be generated: one is
generated by IOPRA (called the IOPRA directed binomial networks), whose nodes
have the same in-degree and out-degree; the other, created by LRA (called the
LRA directed binomial networks), has the same in- and out- degree
distribution, while allowing the in- and out- degree of any node to be different.

\section{Spectral properties in directed networks\label{Sec_spectral}}

\subsection{Spectral radius of directed networks \label{Sec_spectral_radius}}

\begin{figure}
\centering
%EndExpansion
$%
\begin{array}
[c]{cc}%
%TCIMACRO{\FRAME{itbpFU}{3.0511in}{1.6457in}{0.8752in}{\Qcb{(a)}}{}%
%{er1.eps}{\special{ language "Scientific Word";  type "GRAPHIC";
%display "USEDEF";  valid_file "F";  width 3.0511in;  height 1.6457in;
%depth 0.8752in;  original-width 8.9015in;  original-height 4.9882in;
%cropleft "0";  croptop "1";  cropright "1";  cropbottom "0";
%filename 'ER1.eps';file-properties "XNPEU";}}}%
%BeginExpansion
\raisebox{-0.8752in}{\parbox[b]{3.0511in}{\begin{center}
\includegraphics[
height=1.6457in,
width=3.0511in
]%
{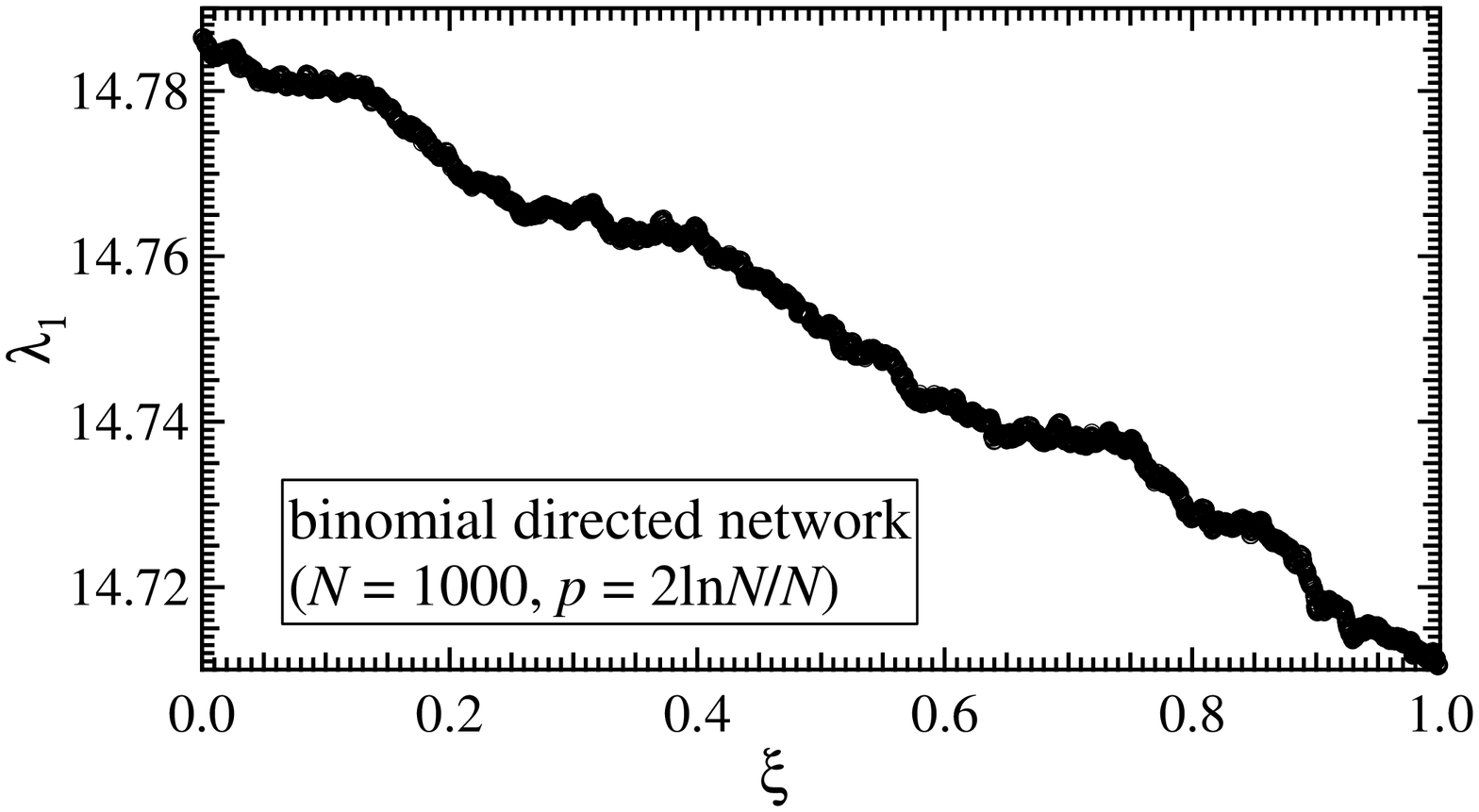}%
\\
(a)
\end{center}}}%
%EndExpansion
&
%TCIMACRO{\FRAME{itbpFU}{3.2024in}{1.6362in}{0.8648in}{\Qcb{(b)}}{}%
%{ba1.eps}{\special{ language "Scientific Word";  type "GRAPHIC";
%display "USEDEF";  valid_file "F";  width 3.2024in;  height 1.6362in;
%depth 0.8648in;  original-width 9.2907in;  original-height 4.9882in;
%cropleft "0";  croptop "1";  cropright "1";  cropbottom "0";
%filename 'BA1.eps';file-properties "XNPEU";}}}%
%BeginExpansion
\raisebox{-0.8648in}{\parbox[b]{3.2024in}{\begin{center}
\includegraphics[
height=1.6362in,
width=3.2024in
]%
{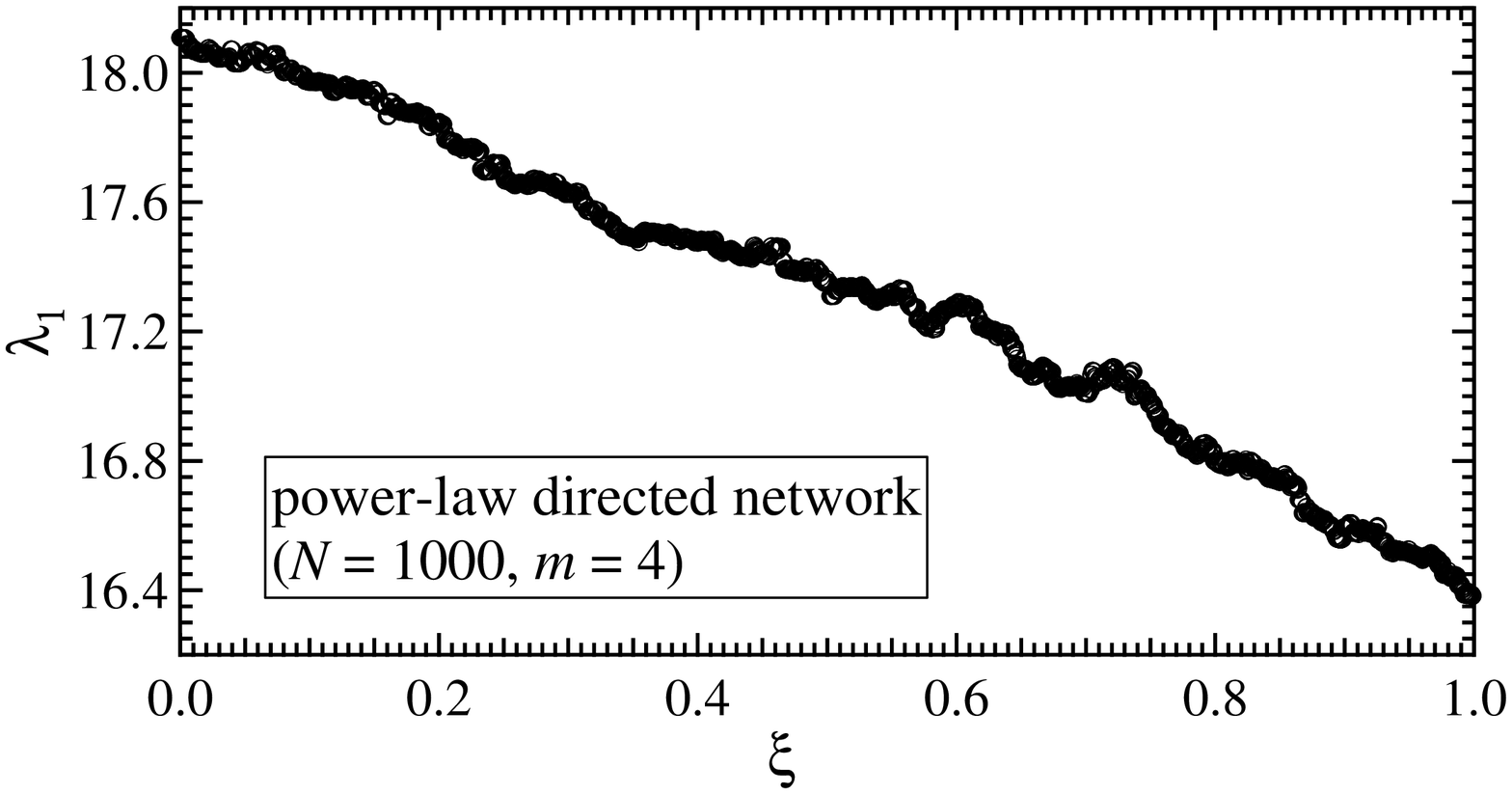}%
\\
(b)
\end{center}}}%
%EndExpansion
\\%
%TCIMACRO{\FRAME{itbpFU}{3.0511in}{1.5843in}{0in}{\Qcb{(c)}}{}{er1ass.eps}%
%{\special{ language "Scientific Word";  type "GRAPHIC";  display "USEDEF";
%valid_file "F";  width 3.0511in;  height 1.5843in;  depth 0in;
%original-width 8.9015in;  original-height 4.9882in;  cropleft "0";
%filename 'ER1ass.eps';file-properties "XNPEU";}}}%
%BeginExpansion
{\parbox[b]{3.0511in}{\begin{center}
\includegraphics[
height=1.5843in,
width=3.0511in
]%
{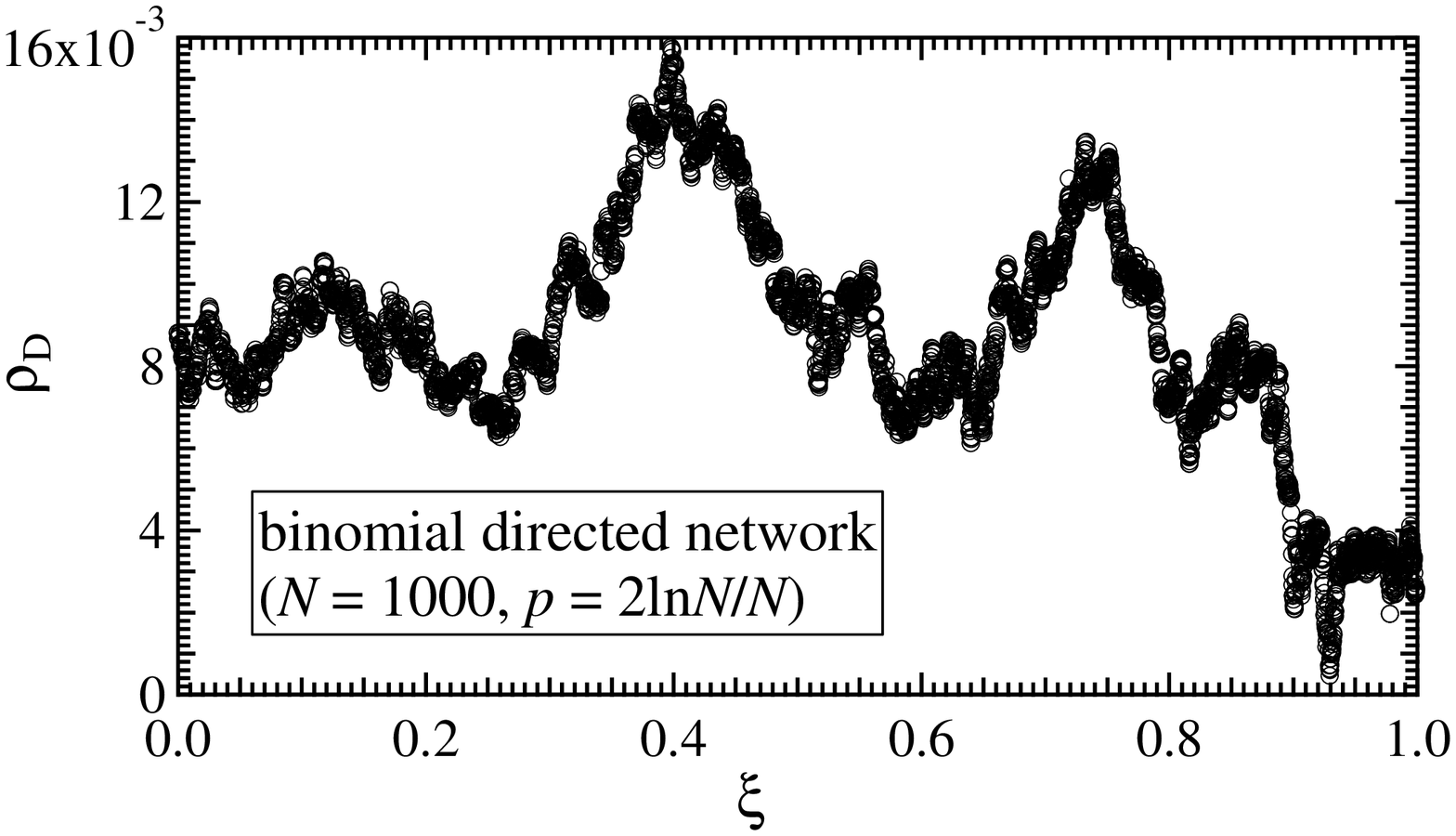}%
\\
(c)
\end{center}}}%
%EndExpansion
&
%TCIMACRO{\FRAME{itbpFU}{3.2197in}{1.5843in}{0.0311in}{\Qcb{(d)}}{}%
%{ba1ass.eps}{\special{ language "Scientific Word";  type "GRAPHIC";
%display "USEDEF";  valid_file "F";  width 3.2197in;  height 1.5843in;
%depth 0.0311in;  original-width 9.3184in;  original-height 4.9882in;
%cropleft "0";  croptop "1";  cropright "1";  cropbottom "0";
%filename 'BA1ass.eps';file-properties "XNPEU";}}}%
%BeginExpansion
\raisebox{-0.0311in}{\parbox[b]{3.2197in}{\begin{center}
\includegraphics[
height=1.5843in,
width=3.2197in
]%
{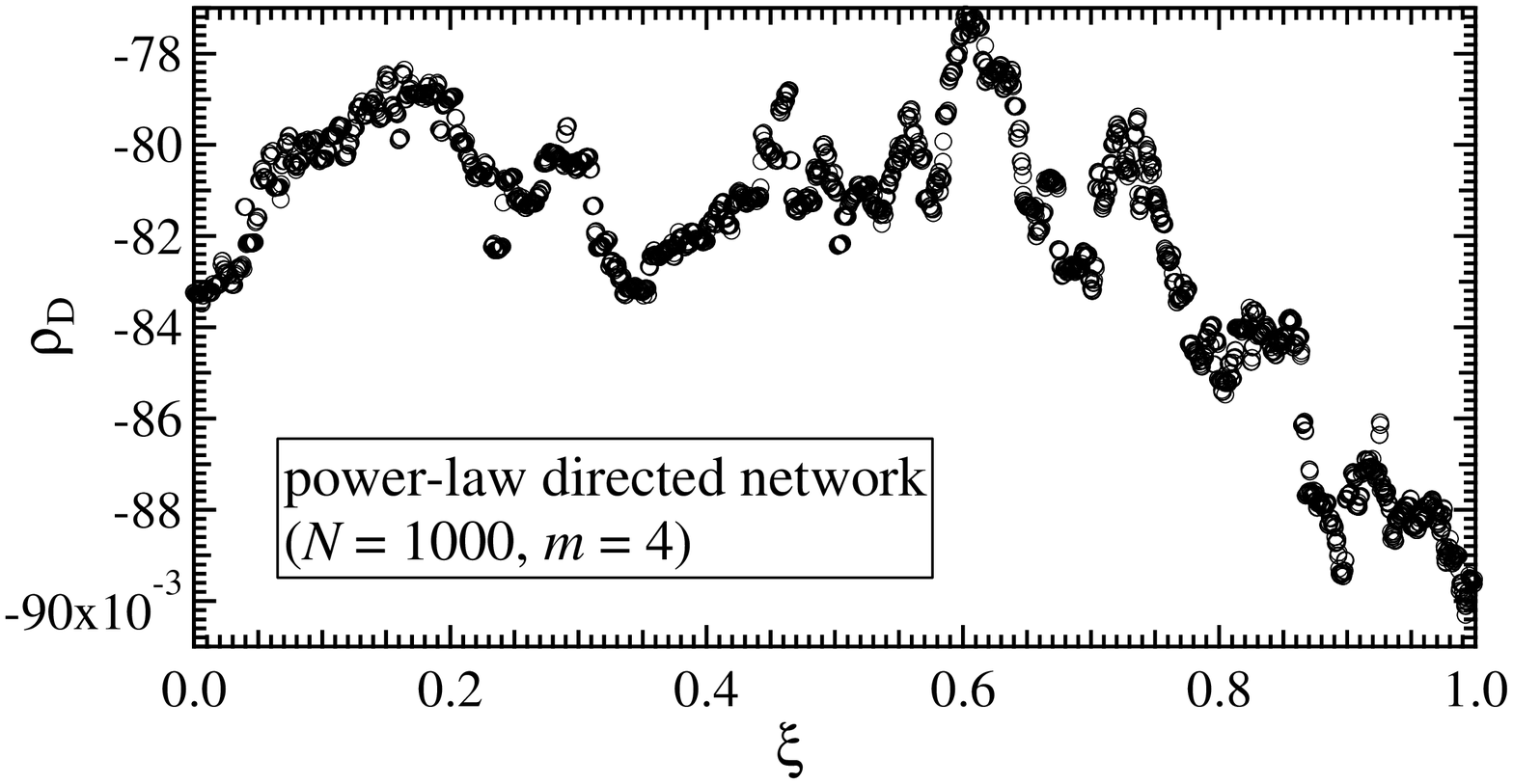}%
\\
(d)
\end{center}}}%
%EndExpansion
\end{array}
$%
%TCIMACRO{\TeXButton{figure_1}{\caption
%{Plot of the spectral radius versus the directionality in (a) and (b), as well as the assortativity versus the directionality in (c) and (d), in both directed binomial and power-law networks generated by IOPRA.}%
%\label{spectral_figure1}
%\end{figure}}}%
%BeginExpansion
\caption
{Plot of the spectral radius versus the directionality in (a) and (b), as well as the assortativity versus the directionality in (c) and (d), in both directed binomial and power-law networks generated by IOPRA.}%
\label{spectral_figure1}
\end{figure}

The adjacency matrix of a directed network is an asymmetric matrix, whose
spectral radius $\lambda_{1}$ is still real by the Perron-Frobenius Theorem
(see \cite{Van Mieghem 2011}). We generate directed networks, with the
directionality $\xi$\ ranging from $0$\ to $1$, by applying IOPRA to
the\textbf{\ }ER ($N=1000$, $p=2$ln$N/N$) and the BA ($N=1000$, $m=4$)
networks gradually. An ER network is connected, if $p>p_{c}\approx$ ln$N/N$
for large $N$, where $p_{c}$ is the disconnectivity threshold. In this work,
we choose $p\geq2p_{c}$ to be sure that the original networks are connected.
The influence of the directionality $\xi$ on the spectral radius $\lambda_{1}$
and the assortativity $\rho_{D}$ is studied in both directed power-law
networks and directed binomial networks (see Figure \ref{spectral_figure1}).%

%TCIMACRO{\TeXButton{figure_1}{\begin{figure}
%\centering}}%
%BeginExpansion
%EndExpansion

\begin{figure}[h]%
\centering
\includegraphics[
height=1.7729in,
width=4.0404in
]%
{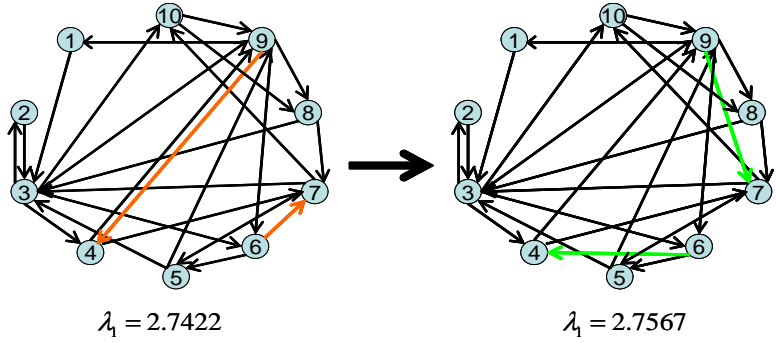}%
\caption{(Color online) Example: the spectral radius increases with the
directionality $\xi$, because of the increase of the assorativity (where
$\rho_{D}(G_{left})=-0.6190$, $\xi(G_{left})=0.8333$, and $\rho_{D}%
(G_{right})=-0.5714$, $\xi(G_{right})=0.9167$).}%
\label{picture2}%
\end{figure}

Apart from some wobbles, the spectral radius $\lambda_{1}$ decreases almost
linearly with the directionality $\xi$. The same phenomenon can also be
observed in large, sparse directed networks (see Appendix D). Moreover, the
assortativity $\rho_{D}$ of the network fluctuates slightly around $0$. We
also have observed a similar phenomenon in large sparse networks. We observe
that the tiny leaps of spectral radius $\lambda_{1}$ happen when the
assortativity $\rho_{D}$ has a rise, which is understandable, because it has
been shown in \cite{Van Mieghem} that the spectral radius $\lambda_{1}$
increases with the increase of the assortativity $\rho_{D}$. Figure
\ref{picture2} exemplifies that the spectral radius $\lambda_{1}$ may increase
instead of decreasing when the directionality increases due to the
assortativity $\rho_{D}$. We will study the effect of the assortativity
$\rho_{D}$ on the decrease of the spectral radius $\lambda_{1}$ with the
directionality $\xi$ in\textbf{\ }Section \ref{Section_infect_assortativity}.%
%TCIMACRO{\FRAME{fhFU}{4.0404in}{1.7729in}{0pt}{\Qcb{(Color online) Example:
%the spectral radius increases with the directionality $\xi$, because of the
%increase of the assorativity (where $\rho_{D}(G_{left})=-0.6190$,
%$\xi(G_{left})=0.8333$, and $\rho_{D}(G_{right})=-0.5714$, $\xi(G_{right}%
%)=0.9167$).}}{\Qlb{picture2}}{picture3.eps}%
%{\special{ language "Scientific Word";  type "GRAPHIC";  display "USEDEF";
%valid_file "F";  width 4.0404in;  height 1.7729in;  depth 0pt;
%original-width 8.3636in;  original-height 3.7239in;  cropleft "0";
%filename 'Picture3.eps';file-properties "XNPEU";}} }%
%BeginExpansion
%EndExpansion

\begin{figure}[ptb]%
\centering
\includegraphics[
height=2.0427in,
width=4.0638in
]%
{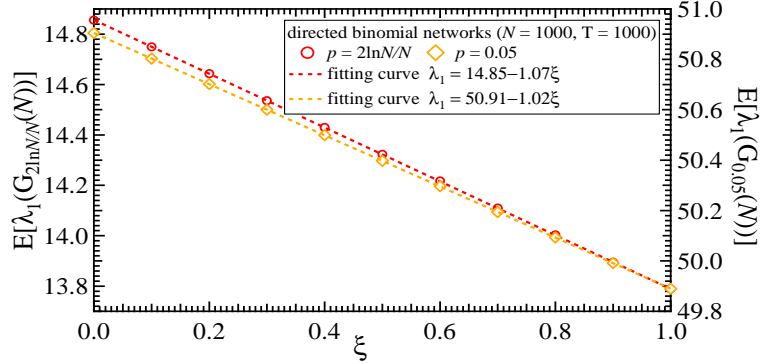}%
\caption{(Color online) Average spectral radius as a function of the
directionality for directed binomial networks generated by LRA with size
$N=1000$. Two values for the link density $p$ are shown: $p=2lnN/N$ (red
circles) and $p=0.05$ (orange diamonds). }%
\label{LRA}%
\end{figure}

With LRA, we generate directed binomial networks with directionality $\xi$
from $0$ to $1$ with step $0.1$. The assortativity $\rho_{D}$ of all the
directed binomial networks generated by LRA is around $0$. Hence, the effect
of the assortativity $\rho_{D}$\textbf{\ }can be ignored here. The spectral
radius $\lambda_{1}$ is calculated in directed networks with different
directionality $\xi$. We performed all the simulations for $10^{3}$ network
realizations. The spectral radius $\lambda_{1}$ is plotted as a function of
the directionality $\xi$ for directed binomial networks with $p=2lnN/N$ and
$p=0.05$ in Figure \ref{LRA}.\
%TCIMACRO{\FRAME{ftbpFU}{4.0638in}{2.0427in}{0pt}{\Qcb{(Color online) Average
%spectral radius as a function of the directionality for directed binomial
%networks generated by LRA with size $N=1000$. Two values for the link density
%$p$ are shown: $p=2lnN/N$ (red circles) and $p=0.05$ (orange diamonds). }%
%}{\Qlb{LRA}}{methodlra.eps}{\special{ language "Scientific Word";
%type "GRAPHIC";  display "USEDEF";  valid_file "F";  width 4.0638in;
%height 2.0427in;  depth 0pt;  original-width 9.2353in;
%original-height 4.9882in;  cropleft "0";  croptop "1";  cropright "1";
%cropbottom "0";  filename 'methodLRA.eps';file-properties "XNPEU";}} }%
%BeginExpansion
%EndExpansion
From the observation, the spectral radius $\lambda_{1}$ is inversely
proportional to the directionality $\xi$ with the factor $\simeq$ $-1$, which
is independent from the link density $p$ of the networks. This observation can
be explained by the following proposition.

\begin{proposition}
Let $G^{(\xi=0)}=G_{p}(N)$ be a connected Erd\H{o}s-R\'{e}nyi (ER) random
graph with a finite $N$, and let $G^{(\xi)}$ be a directed binomial network
generated by LRA whose in- and out-degree follow the same binomial
distribution as\textbf{\ }$G_{p}(N)$. The average spectral radius
satisfies\label{proposition}
\begin{equation}
E[\lambda_{1}(G^{(\xi)})]\simeq E[\lambda_{1}(G_{p}(N))]-\xi
\label{spectral_decrease}%
\end{equation}

\end{proposition}

\textbf{Arguments:}

\begin{figure}[h]%
\centering
\includegraphics[
height=2.1672in,
width=4.1978in
]%
{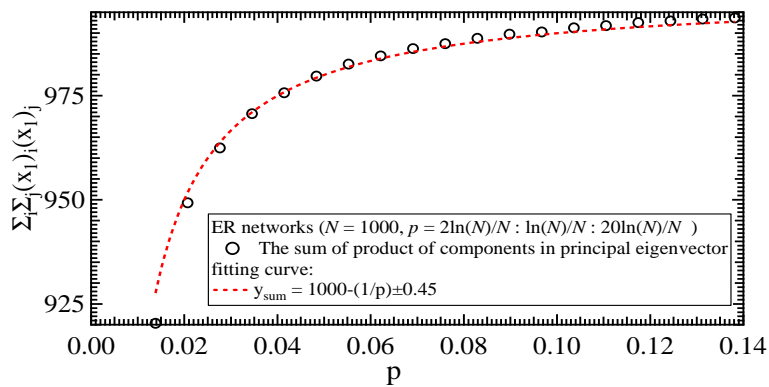}%
\caption{(Color online) Sum of the product of components in the principal
eigenvector as a function of the link density $p$ in ER networks ($N=1000$).
\ }%
\label{sumproduct}%
\end{figure}

A directed binomial network $G^{(\xi)}$ generated by LRA with link density
$p$, can be equivalently constructed by randomly adding $2p\xi(_{2}^{N})$
unidirectional links to a bidirectional ER network $G_{p(1-\xi)}(N)$ with size
$N$ and link density $p(1-\xi)$. The average spectral radius \cite[pp. 173,
art. 137]{Van Mieghem 2011} of $G_{p(1-\xi)}(N)$ is
\[
E[\lambda_{1}(G_{p(1-\xi)}(N))]=(N-2)p(1-\xi)+1+O(\frac{1}{\sqrt{N}})\text{.}%
\]
The principal eigenvector of an adjacency matrix $A$ is denoted by $x_{1}$
obeying the normalization $x_{1}^{T}x_{1}=1$.\ Let $C$ denote the adjacency
matrix of the resulting network after adding one unidirectional link to
network $G$. The largest eigenvalue is increased due to the addition of the
link ($i\rightarrow j$) \cite[pp. 236, Lemma 7]{Van Mieghem 2011} as
\[
\lambda_{1}(C)\simeq\lambda_{1}(A)+(x_{1})_{i}(x_{1})_{j}\text{,}%
\]
where the increase is strict if the adjacency matrix $A$ is irreducible.
Hence, the average increase of the spectral radius by adding $m$
unidirectional links in random networks is obtained as%
\[
E[\lambda_{1}(C)-\lambda_{1}(A)]\simeq mE[(x_{1})_{i}(x_{1})_{j}]\text{.}%
\]
The sum of the product of components in the principal eigenvector of
Erd\H{o}s-R\'{e}nyi networks is approximated by a function of link density $p$
(see Figure \ref{sumproduct}). The fitting function can be expressed as,%
\[
E\left[  \sum_{j=1}^{N}\sum_{i=1}^{N}(x_{1})_{i}(x_{1})_{j}\right]
=N-\frac{1}{p}+O(1)\text{,}%
\]
when the network is connected. Since $x_{1}^{T}x_{1}=1$ and since the
expectation $E[.]$ is a linear operator, we obtain%

\begin{equation}
E\left[  (x_{1})_{i}(x_{1})_{j}\right]  =\frac{N-\frac{1}{p}-1}{N(N-1)}%
+O(\frac{1}{N^{2}})\text{,} \label{product}%
\end{equation}
when $i\neq j$. Directed binomial networks generated by LRA from ER with $N$
and $p$, have the same $E\left[  (x_{1})_{i}(x_{1})_{j}\right]  $. Hence, the
average spectral radius of the directed network obtained by adding
$m=2p\xi(_{2}^{N})$ unidirectional links to\ the network $G_{p(1-\xi)}(N)$ can
be approximated by
\[
E[\lambda_{1}(G^{_{(\xi)}})]\simeq E[\lambda_{1}(G_{p(1-\xi)}(N))]+2\left(
N(N-1)/2\right)  p\xi E\left[  (x_{1})_{i}(x_{1})_{j}\right]  \text{.}%
\]
Using (\ref{product}),%

\[
E[\lambda_{1}(G^{_{(\xi)}})]\simeq(N-2)p+1-\xi+O(\frac{1}{\sqrt{N}})\text{,}%
\]
which leads to (1).
$\ \ \ \ \ \ \ \ \ \ \ \ \ \ \ \ \ \ \ \ \ \ \ \ \ \ \ \ \ \ \ \ \ \ \ \ \ \ \ \ \ \ \ \ \ \ \ \ \ \ \ \ \ \ \ \ \ \ \ \ \ \ \ \ \ \ \ \ \ \ \ \ \ \ \ \ \ \ \ \ \ \ \ \ \ \ \ \ \ \ \ \ \ \ \ \ \ \ \ \ \ \ \ \ \ \ \ \ \ \ \square
$

Juh\'{a}sz \cite{Juhasz} also pointed out that the largest eigenvalue
$\lambda_{1}(G^{(\xi=1)})$ of a directed random network with link density $p$
and size $N$ is almost surely $Np$, when $N$ is large. In ER random
networks\textbf{, }the spectral radius $E[\lambda_{1}(G^{(\xi=0)})]\rightarrow
Np+1$, when $N$ is large (see \cite[pp. 173, art. 137]{Van Mieghem 2011}).
Both earlier results are consistent with Proposition \ref{proposition}, and
support that the proportionality factor between the spectral radius
$\lambda_{1}$\ and the directionality $\xi$\ is around $-1$.%

%TCIMACRO{\FRAME{fhFU}{4.1978in}{2.1672in}{0pt}{\Qcb{(Color online) Sum of the
%product of components in the principal eigenvector as a function of the link
%density $p$ in ER networks ($N=1000$). \ }}{\Qlb{sumproduct}}{sumproduct.eps}%
%{\special{ language "Scientific Word";  type "GRAPHIC";  display "USEDEF";
%valid_file "F";  width 4.1978in;  height 2.1672in;  depth 0pt;
%original-width 9.0278in;  original-height 4.9882in;  cropleft "0";
%filename 'sumproduct.eps';file-properties "XNPEU";}} }%
%BeginExpansion
%EndExpansion

Proposition \ref{proposition} also reveals the effect of the size $N$ on the
relative largest decrease of the spectral radius $\Lambda=\frac{\lambda
_{1}(G^{(\xi=0)})-\lambda_{1}(G^{(\xi=1)})}{\lambda_{1}(G^{(\xi=0)})}$. We
predict that $\Lambda\rightarrow0$ if $N\rightarrow\infty$ for directed
binomial networks, because the decrease of the spectral radius ($\lambda
_{1}(G^{(\xi=0)})-\lambda_{1}(G^{(\xi=1)})$) is almost a constant value,
whereas the spectral radius $\lambda_{1}(G^{(\xi=0)})$ of dense directed
binomial networks increases with the size of the networks. This implies that
the effect of the directionality $\xi$ on the spectral radius is small in
large dense binomial networks.

\subsection{Principal eigenvector in directed
networks\label{Section_eigenvector}}

\begin{figure}[h]%
\centering
\includegraphics[
height=2.0877in,
width=4.2047in
]%
{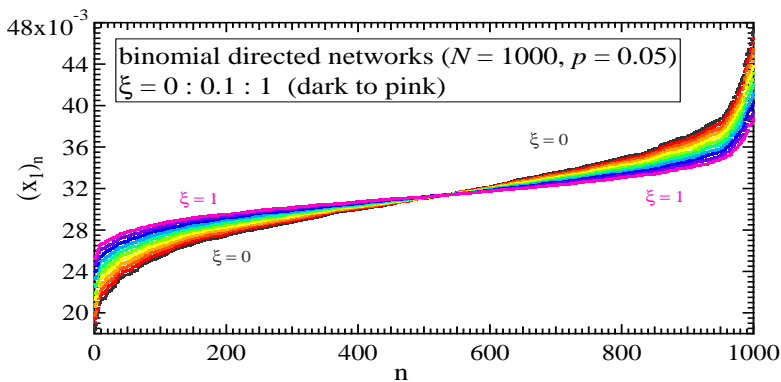}%
\caption{(Color online) Change of the components of principal eigenvector from
bidirectional networks to directed networks. }%
\label{Eigenvector}%
\end{figure}

\begin{figure}[b]
\centering
%EndExpansion
$%
\begin{array}
[c]{cc}%
%TCIMACRO{\FRAME{itbpFU}{3.1618in}{1.7045in}{-0.0104in}{\Qcb{(a)}}{}%
%{varer.eps}{\special{ language "Scientific Word";  type "GRAPHIC";
%display "USEDEF";  valid_file "F";  width 3.1618in;  height 1.7045in;
%depth -0.0104in;  original-width 8.7121in;  original-height 4.7971in;
%cropleft "0";  croptop "1";  cropright "1";  cropbottom "0";
%filename 'VarER.eps';file-properties "XNPEU";}} }%
%BeginExpansion
\raisebox{0.0104in}{\parbox[b]{3.1618in}{\begin{center}
\includegraphics[
height=1.7045in,
width=3.1618in
]%
{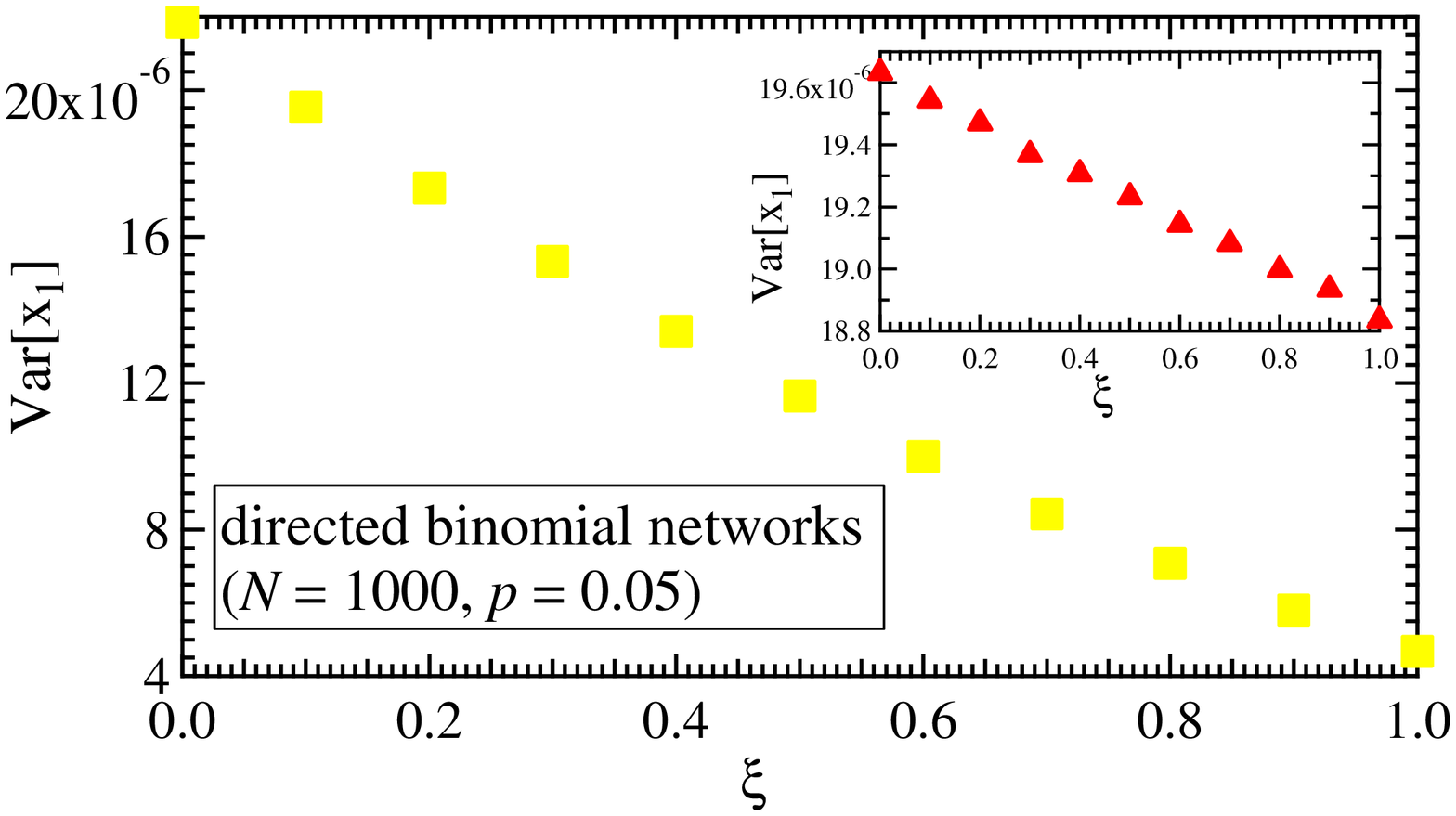}%
\\
(a)
\end{center}}}
%EndExpansion
&
%TCIMACRO{\FRAME{itbpFU}{3.3114in}{1.7193in}{0.0104in}{\Qcb{(b)}}%
%{}{Varba10times.eps}{\special{ language "Scientific Word";  type "GRAPHIC";
%display "USEDEF";  valid_file "F";  width 3.3114in;  height 1.7193in;
%depth 0.0104in;  original-width 8.9015in;  original-height 4.9882in;
%cropleft "0";  croptop "1";  cropright "1";  cropbottom "0";
%filename 'VArBA10times.eps';file-properties "XNPEU";}} }%
%BeginExpansion
\raisebox{-0.0104in}{\parbox[b]{3.3114in}{\begin{center}
\includegraphics[
height=1.7193in,
width=3.3114in
]%
{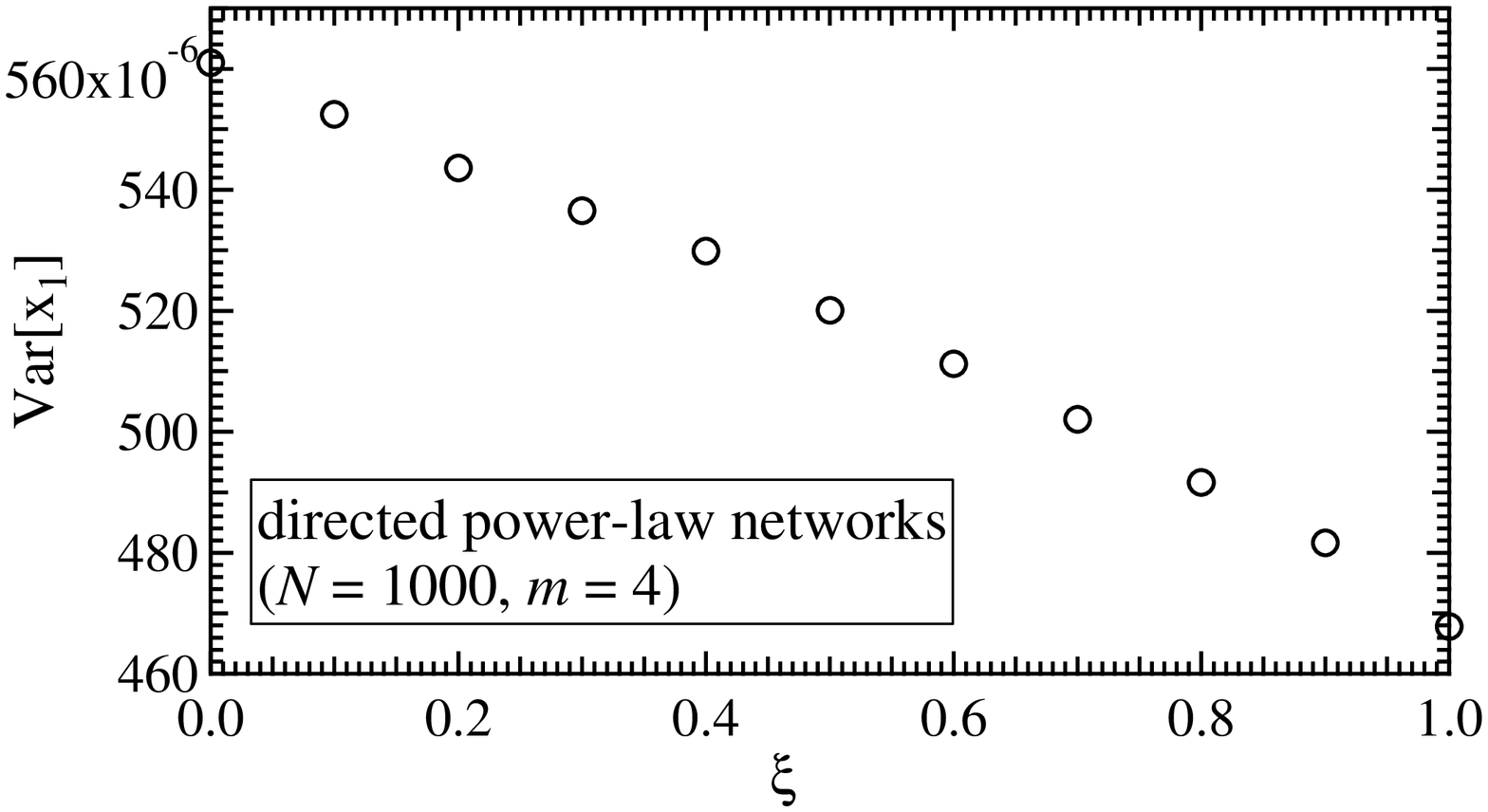}%
\\
(b)
\end{center}}}
%EndExpansion
\end{array}
$%
%TCIMACRO{\TeXButton{figure_3}{\caption
%{(Color online) Plot of the variance of the principal eigenvector versus the directionality (a) in directed binomial networks (generated by LRA in yellow squares and by IOPRA in red triangles) and (b) in directed power-law networks by IOPRA ($10^{3}%
%$ network realizations).}\label{spectral_figure3}
%\end{figure}}}%
%BeginExpansion
\caption
{(Color online) Plot of the variance of the principal eigenvector versus the directionality (a) in directed binomial networks (generated by LRA in yellow squares and by IOPRA in red triangles) and (b) in directed power-law networks by IOPRA ($10^{3}%
$ network realizations).}\label{spectral_figure3}
\end{figure}

The principal eigenvector $x_{1}$ was first proposed as a centrality metric by
Bonacich \cite{Bonacich} in $1987$, to indicate the influence of each node.
For example, the decrease of the spectral radius \cite{Van Mighem_removelinks,
CLi_Removingnodes} by removing nodes, can be characterized by the
corresponding principal eigenvector components. In this section, we explore
the principal eigenvector in directed networks. The principal eigenvector of
the directed networks, with the directionality $\xi$ from $0$ to $1$ with step
$0.1$, are calculated. Then, the components of the principal eigenvector are
sorted in an ascending order. For each $\xi$, we simulate $10^{3}$\ network
realizations and compute the average sorted principal eigenvector components.
Figure \ref{Eigenvector} illustrates that the components of the principal
eigenvector are more uniform in directed binomial networks: the principal
eigenvector $x_{1}\rightarrow\frac{u}{\sqrt{N}}$ as $\xi\rightarrow1$;
moreover, the variance of components of the principal eigenvector linearly
decreases with the directionality $\xi$ in both directed binomial networks and
the directed power-law networks (see Figure \ref{spectral_figure3}). The
observation implies that when the directionality is larger, the influence of
each node on the spectral radius is more similar. This experimental evidence
suggests that increasing the directionality enables all nodes to contribute
more similarly to the robustness against epidemic in directed networks.

The decrease of the variance $Var[x_{1}]$ in directed binomial networks by LRA
is larger than that in directed binomial networks by IOPRA (see Figure
\ref{spectral_figure3}(a)). The connections in LRA directed binomial networks
are more random than that in IOPRA directed binomial networks, in the sense
that LRA allows each node to have a different in- and out- degree, although
the in- and out- degree distribution are the same in both LRA\ and IOPRA
binomial networks. As a consequence, the principal eigenvector $x_{\mathbf{1}%
}$\ is more uniform with a smaller $Var[x_{1}]$\ in LRA directed binomial
networks than that in IOPRA directed binomial networks when the directionality
is the same. Thus, nodes in LRA directed binomial networks have more equal
contributions to the spectral radius than nodes in IOPRA directed binomial
networks with the same directionality. Li \textit{et al.}
\cite{CLi_Networking} have shown that both a large variance of the degree and
a large assortativity $\rho_{D}$ contribute to a large variance $Var[x_{1}]$
of the components of the principal eigenvector $x_{1}$. Here, we point out
further that a large directionality $\xi$\ leads to a small variance
$Var[x_{1}]$\ of the components of $x_{1}$.%
%TCIMACRO{\FRAME{fhFU}{4.2047in}{2.0877in}{0pt}{\Qcb{(Color online) Change of
%the components of principal eigenvector from bidirectional networks to
%directed networks. }}{\Qlb{Eigenvector}}{principaleigenvector.eps}%
%{\special{ language "Scientific Word";  type "GRAPHIC";  display "USEDEF";
%valid_file "F";  width 4.2047in;  height 2.0877in;  depth 0pt;
%original-width 9.6669in;  original-height 5.5287in;  cropleft "0";
%filename 'principaleigenvector.eps';file-properties "XNPEU";}} }%
%BeginExpansion
%EndExpansion
%

%TCIMACRO{\TeXButton{figure_3}{\begin{figure}[b]
%\centering}}%
%BeginExpansion
%EndExpansion

\subsection{Spectral gap of directed networks\label{Sec_spectral gap}}

\begin{figure}
\centering
%EndExpansion
$%
\begin{array}
[c]{cc}%
%TCIMACRO{\FRAME{itbpFU}{3.1868in}{1.8031in}{0in}{\Qcb{(a)}}{}%
%{spectralgap1000.eps}{\special{ language "Scientific Word";  type "GRAPHIC";
%display "USEDEF";  valid_file "F";  width 3.1868in;  height 1.8031in;
%depth 0in;  original-width 8.9015in;  original-height 4.9882in;
%cropleft "0";  croptop "1";  cropright "1";  cropbottom "0";
%filename 'spectralgap1000.eps';file-properties "XNPEU";}} }%
%BeginExpansion
{\parbox[b]{3.1868in}{\begin{center}
\includegraphics[
height=1.8031in,
width=3.1868in
]%
{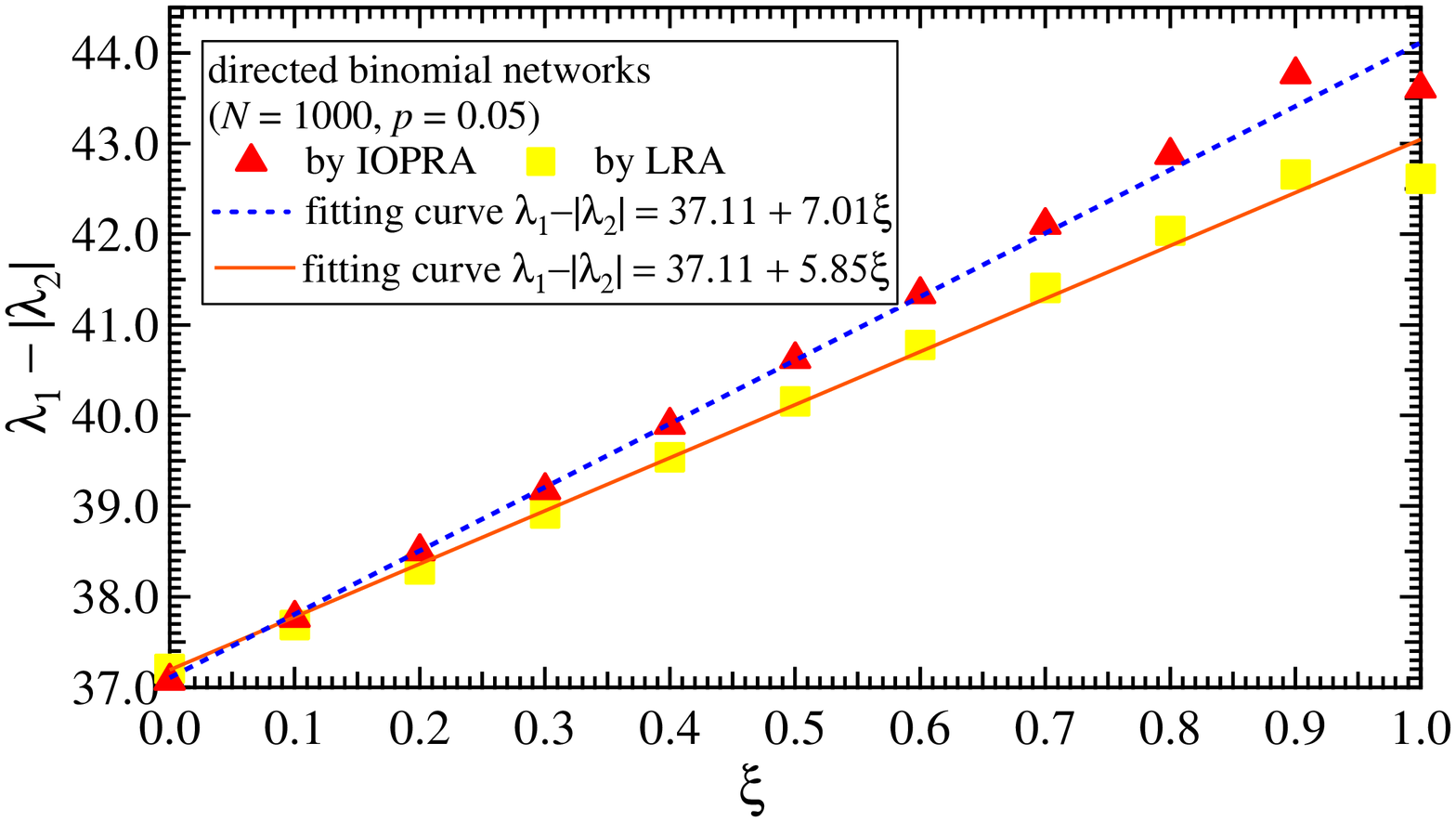}%
\\
(a)
\end{center}}}
%EndExpansion
&
%TCIMACRO{\FRAME{itbpFU}{3.2379in}{1.8031in}{0in}{\Qcb{(b)}}{}%
%{spectralgapba1000.eps}{\special{ language "Scientific Word";
%type "GRAPHIC";  display "USEDEF";  valid_file "F";  width 3.2379in;
%height 1.8031in;  depth 0in;  original-width 8.777in;
%original-height 4.9882in;  cropleft "0";  croptop "1";  cropright "1";
%cropbottom "0";  filename 'spectralgapBA1000.eps';file-properties "XNPEU";}}
%}%
%BeginExpansion
{\parbox[b]{3.2379in}{\begin{center}
\includegraphics[
height=1.8031in,
width=3.2379in
]%
{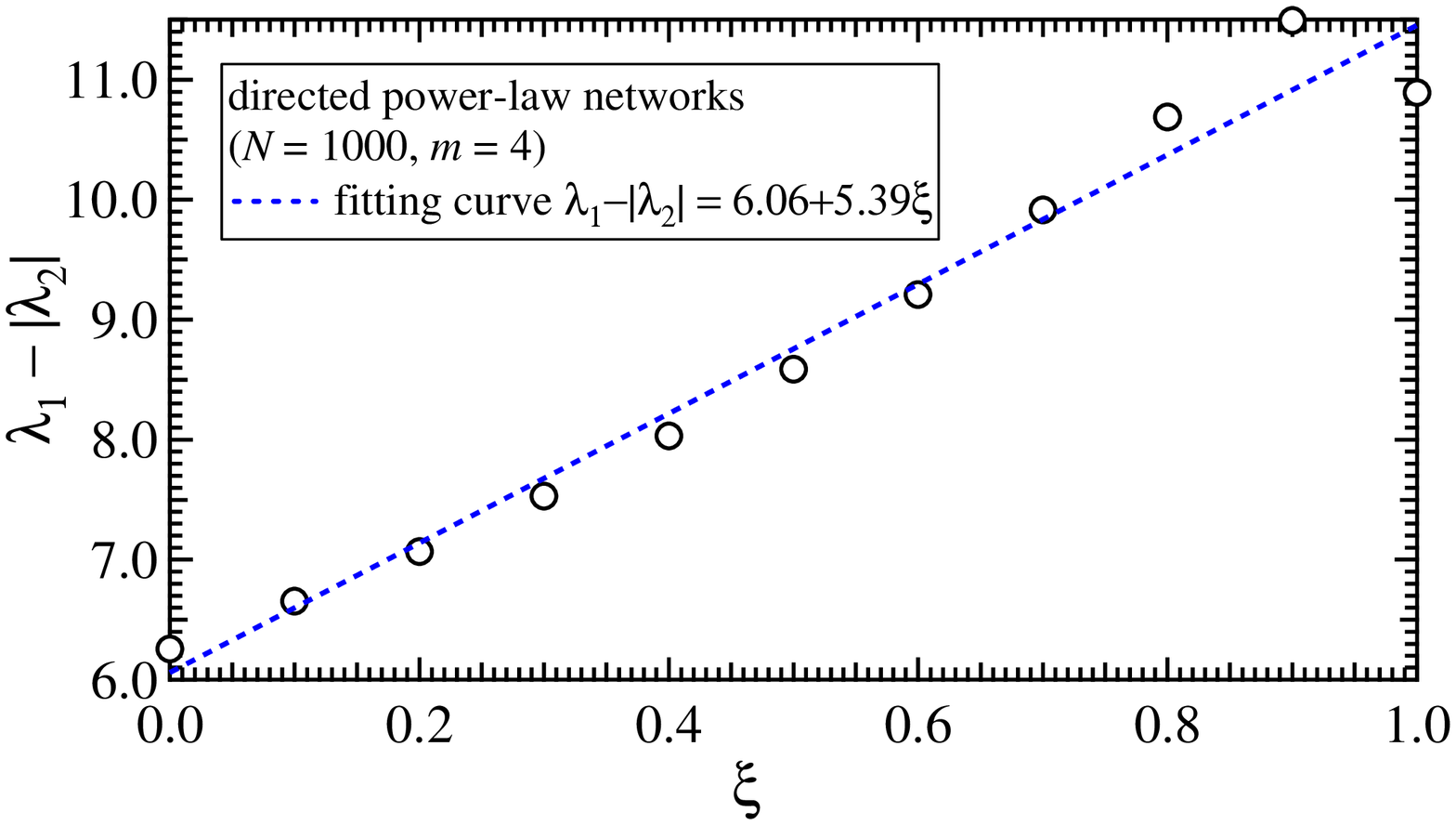}%
\\
(b)
\end{center}}}
%EndExpansion
\end{array}
$%
%TCIMACRO{\TeXButton{figure_4}{\caption
%{(Color online) Plot of the spectral gap as a function of the directionality (A) in directed binomial networks (generated by LRA in yellow square and by IOPRA in red triangle) and (B) in directed power-law networks ($10^{3}%
%$ network realizations).}\label{spectral_figure4}
%\end{figure}}}%
%BeginExpansion
\caption
{(Color online) Plot of the spectral gap as a function of the directionality (A) in directed binomial networks (generated by LRA in yellow square and by IOPRA in red triangle) and (B) in directed power-law networks ($10^{3}%
$ network realizations).}\label{spectral_figure4}
\end{figure}

The difference $(\lambda_{1}-\lambda_{2})$ between the largest eigenvalue
$\lambda_{1}$ and the second largest eigenvalue $\lambda_{2}$ is called the
spectral gap. All eigenvalues of the symmetric adjacency matrix of an
undirected network are real. Here we focus on the directed networks, whose
adjacency matrix is asymmetric. The eigenvalues of directed networks can be
complex numbers (as exemplified in Figure \ref{Eigenvalues} in Appendix B). In
directed networks, the spectral gap is defined as $\lambda_{1}-\left\vert
\lambda_{2}\right\vert $, where $\left\vert \lambda_{2}\right\vert $\ is the
modulus of $\lambda_{2}$. The spectral gap $\lambda_{1}-\left\vert \lambda
_{2}\right\vert $ increases with the directionality $\xi$ in both the directed
binomial networks and the directed power-law networks (see Figure
\ref{spectral_figure4}). As introduced in Sec. \ref{Sec_spectral_radius}, the
spectral radius decreases with the directionality.\textbf{\ }Our observation
implies that the second largest eigenvalue $\left\vert \lambda_{2}\right\vert
$ decreases with the directionality faster than the spectral radius. The
larger the spectral gap is, the faster a random walk converges to its
steady-state \cite[pp. 64]{Van Mieghem 2011}. Thus, the dynamic process in a
directed network reaches the steady-state faster than that in an undirected
network with the same degree distribution. Figure \ref{spectral_figure4} (a)
implies that a dynamic process is slightly faster to reach the steady-state in
IOPRA directed binomial networks than in LRA directed binomial networks. The
existence of large spectral gap together with a uniform degree distribution
results in higher structural sturdiness and robustness against node and link
failures \cite{Thai}. Hence, directed networks with high directionality $\xi
$\ and a uniform degree distribution are more robust than undirected networks
with large variance of degree.%

%TCIMACRO{\TeXButton{figure_4}{\begin{figure}
%\centering}}%
%BeginExpansion
%EndExpansion

\subsection{Algebraic connectivity of directed networks}

\begin{figure}
\centering
%EndExpansion
$%
\begin{array}
[c]{cc}%
%TCIMACRO{\FRAME{itbpFU}{3.1557in}{1.8239in}{-0.0104in}{\Qcb{(a)}}{}%
%{alger.eps}{\special{ language "Scientific Word";  type "GRAPHIC";
%display "USEDEF";  valid_file "F";  width 3.1557in;  height 1.8239in;
%depth -0.0104in;  original-width 8.9015in;  original-height 4.9882in;
%cropleft "0";  croptop "1";  cropright "1";  cropbottom "0";
%filename 'AlgER.eps';file-properties "XNPEU";}} }%
%BeginExpansion
\raisebox{0.0104in}{\parbox[b]{3.1557in}{\begin{center}
\includegraphics[
height=1.8239in,
width=3.1557in
]%
{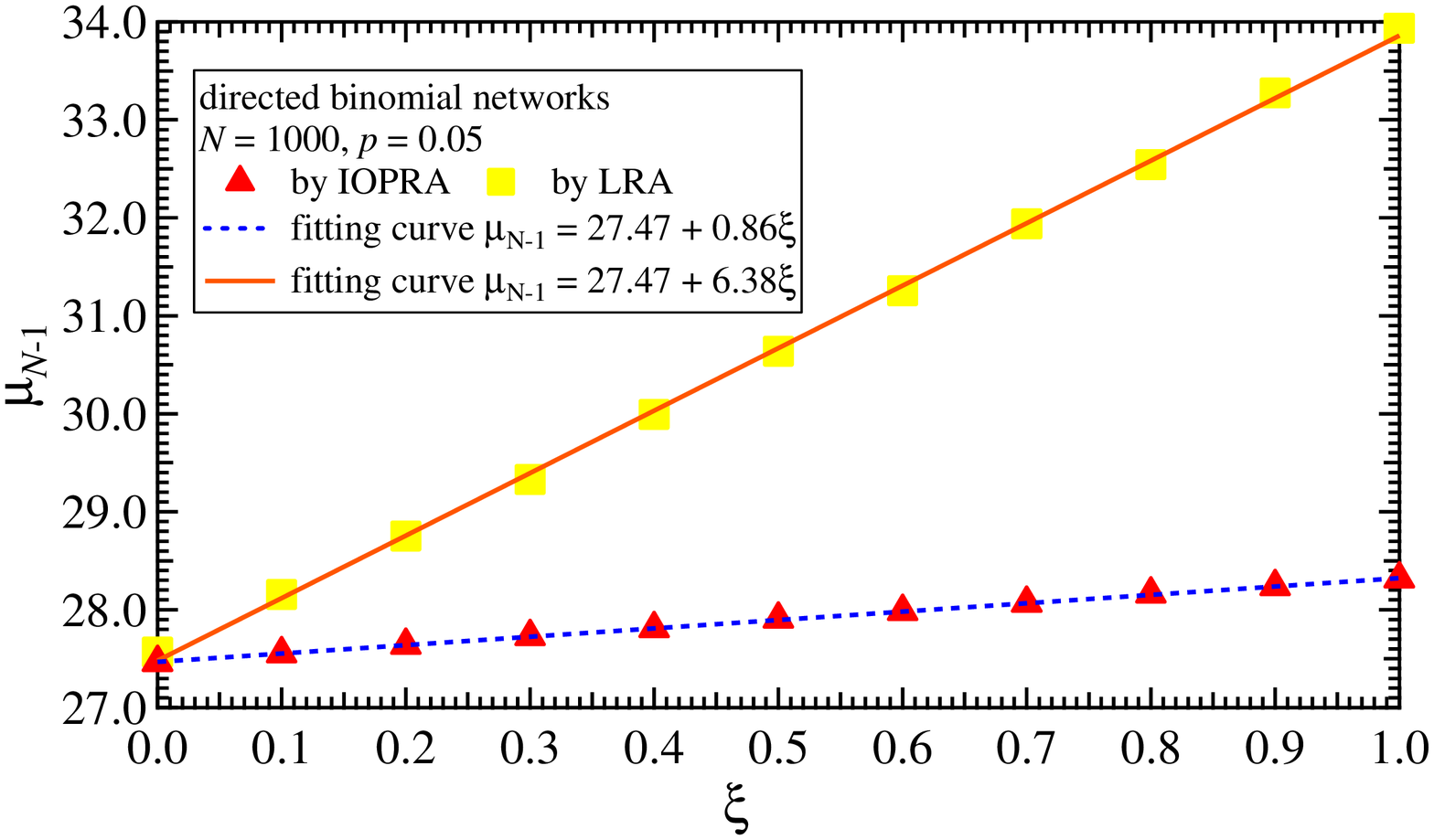}%
\\
(a)
\end{center}}}
%EndExpansion
&
%TCIMACRO{\FRAME{itbpFU}{3.3425in}{1.8222in}{0in}{\Qcb{(b)}}{}{algba.eps}%
%{\special{ language "Scientific Word";  type "GRAPHIC";  display "USEDEF";
%valid_file "F";  width 3.3425in;  height 1.8222in;  depth 0in;
%original-width 8.3636in;  original-height 4.8101in;  cropleft "0";
%filename 'AlgBA.eps';file-properties "XNPEU";}} }%
%BeginExpansion
{\parbox[b]{3.3425in}{\begin{center}
\includegraphics[
height=1.8222in,
width=3.3425in
]%
{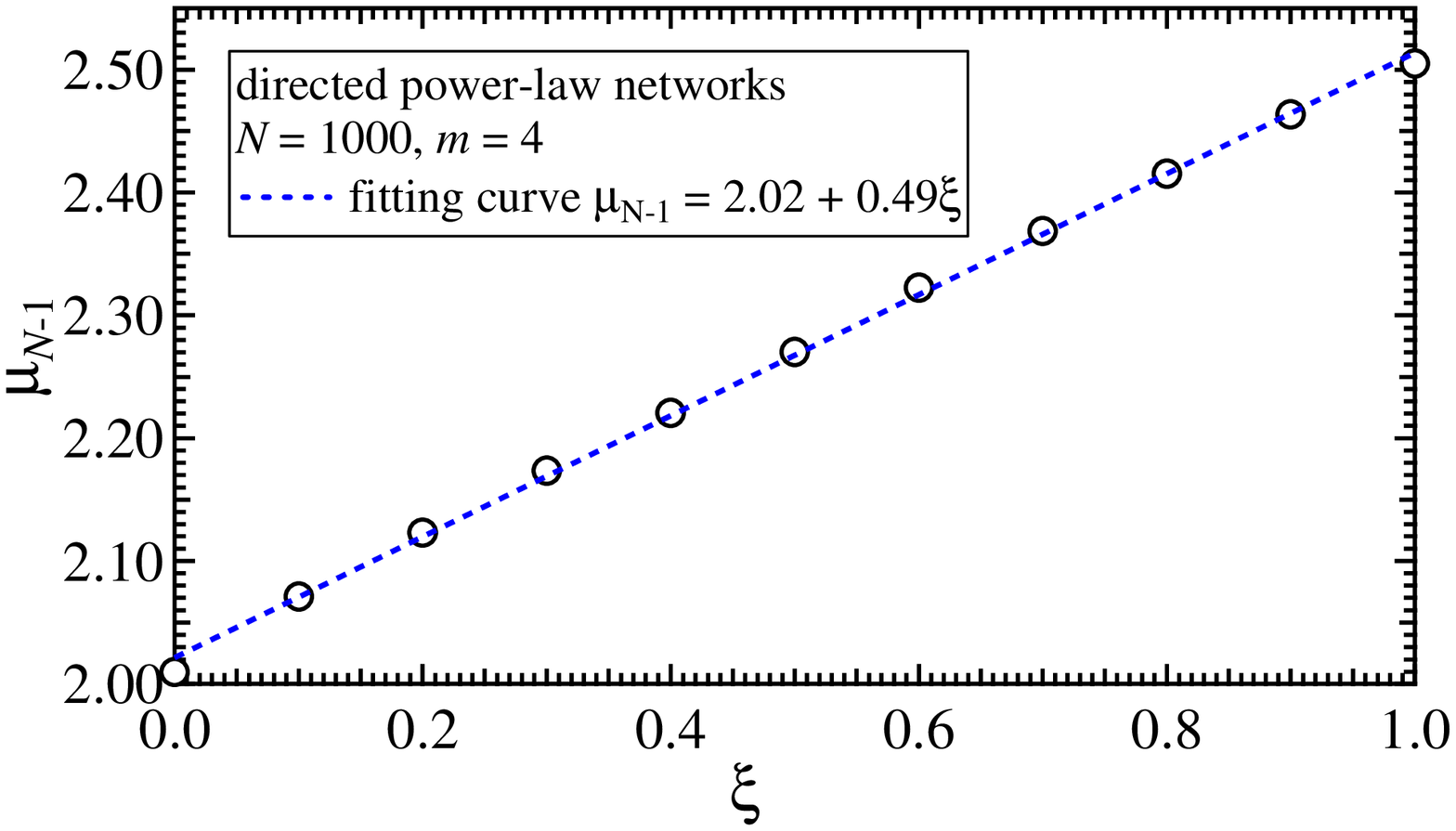}%
\\
(b)
\end{center}}}
%EndExpansion
\end{array}
$%
%TCIMACRO{\TeXButton{figure_5}{\caption
%{(Color online) Plot of the algebraic connectivity as a function of the directionality (A) in directed binomial networks (generated by LRA in yellow square and by IOPRA in red triangle) and (B) in directed power-law networks ($10^{3}%
%$ network realizations).}\label{spectral_figure5}
%\end{figure}}}%
%BeginExpansion
\caption
{(Color online) Plot of the algebraic connectivity as a function of the directionality (A) in directed binomial networks (generated by LRA in yellow square and by IOPRA in red triangle) and (B) in directed power-law networks ($10^{3}%
$ network realizations).}\label{spectral_figure5}
\end{figure}

The Laplacian matrix \cite{Bapat} is defined as $Q=\frac{1}{2}BB^{T}$, where
the incidence matrix $B$ is an $N\times L$ matrix with elements \cite{Van
Mieghem 2011}%
\[
b_{il}=\left\{
\begin{array}
[c]{c}%
1\text{ \ if link }e_{l}=i\rightarrow j\\
-1\text{ if link }e_{l}=j\rightarrow i\\
0\text{ \ \ \ \ \ \ \ \ \ otherwise.}%
\end{array}
\right.
\]
The Laplacian\ matrix can be equivalently expressed as $Q=\Delta-\bar{A}$,
where $\Delta=\frac{1}{2}\left(  \Delta_{in}+\Delta_{out}\right)  $,
$\Delta_{in}$\ and $\Delta_{out}$\ are diagonal matrices which contain the
in-degree and out-degree of each node respectively, and $\bar{A}=\frac{1}%
{2}(A+A^{T})$. If the network is an undirected network, $\bar{A}$ is the
adjacency matrix $A$ and $\Delta=$ \textrm{diag}$(d_{1},d_{2},\cdots,d_{N})$
is the degree matrix. The second smallest eigenvalue $\mu_{N-1}$ of the
Laplacian $Q$ was named algebraic connectivity by Fiedler \cite{Fiedler}. The
Laplacian $Q$\ is always symmetric as defined. Hence, the algebraic
connectivity of a directed and connected network is a positive real number.
The algebraic connectivity, together with the spectral gap, quantifies the
robustness and the network's well-connectedness. The larger the algebraic
connectivity is, the more difficult it is to cut the network into disconnected
parts. Here, we study the influence of the directionality $\xi$\ on the
algebraic connectivity $\mu_{N-1}$ of directed networks. As illustrated in
Figure \ref{spectral_figure5}, the algebraic connectivity increases with the
directionality $\xi$ in both the\ directed binomial networks and the directed
power-law networks. This suggests that the directed networks with high
directionality are more difficult to break into parts and synchronize faster.
As the directionality increases, the number of none-zero elements of $\bar{A}$
increases and the variance of the elements of $\bar{A}$\ decreases. This could
be one possible reason why the network is better connected. Moreover, the
algebraic connectivity $\mu_{N-1}$\ is greater in the LRA directed binomial
networks than in the IOPRA\ binomial networks (see Figure
\ref{spectral_figure5}(a)).

The algebraic connectivity $\mu_{N-1}$ approaches the spectral gap
$\lambda_{1}-\lambda_{2}$, as the network tends to be regular bidirectional
networks \cite[pp. 71]{Van Mieghem 2011}, which suggests the spectral gap is
related to the algebraic connectivity. Figures \ref{spectral_figure4} and
\ref{spectral_figure5} show that both,\ the spectral gap and the algebraic
connectivity, increase with the directionality $\xi$\ in directed networks,
which is consistent with the relation between the algebraic connectivity and
the spectral gap.%

%TCIMACRO{\TeXButton{figure_5}{\begin{figure}
%\centering}}%
%BeginExpansion
%EndExpansion

\section{Effects of the assortativity on the spectral radius of directed
networks\label{Section_infect_assortativity}}

In the directed networks generated by applying IOPRA to ER or BA networks, the
in- and in- degree correlation, the in- and out- degree correlation, the out-
and in- degree correlation and the out- and out- degree correlation are the
same. Thus, the four correlations are all referred as the degree correlation
(or the assortativity). In Section \ref{Sec_spectral}, we have discussed how
the spectral properties change with the directionality in directed networks,
where the assortativity is always close to zero. Here, we study how the
spectral radius $\lambda_{1}$ changes with the directionality $\xi$ when the
assortativity $\rho_{D}$ is the same, and how the change of the spectral
radius $\lambda_{1}$ with $\xi\ $is influenced by the assortativity in
directed networks. Two approaches are applied to investigate this problem.%

%TCIMACRO{\TeXButton{figure_7}{\begin{figure}[b]
%\centering}}%
%BeginExpansion
\begin{figure}[b]
\centering
%EndExpansion
$%
\begin{array}
[c]{cc}%
%TCIMACRO{\FRAME{ihFU}{3.1384in}{1.7608in}{0in}{\Qcb{(a) one network
%realization}}{}{effectofdirection.eps}{\special{ language "Scientific Word";
%type "GRAPHIC";  display "USEDEF";  valid_file "F";  width 3.1384in;
%height 1.7608in;  depth 0in;  original-width 8.8747in;
%original-height 4.9882in;  cropleft "0";  croptop "1";  cropright "1";
%cropbottom "0";  filename 'effectofdirection.eps';file-properties "XNPEU";}}
%}%
%BeginExpansion
{\parbox[b]{3.1384in}{\begin{center}
\includegraphics[
height=1.7608in,
width=3.1384in
]%
{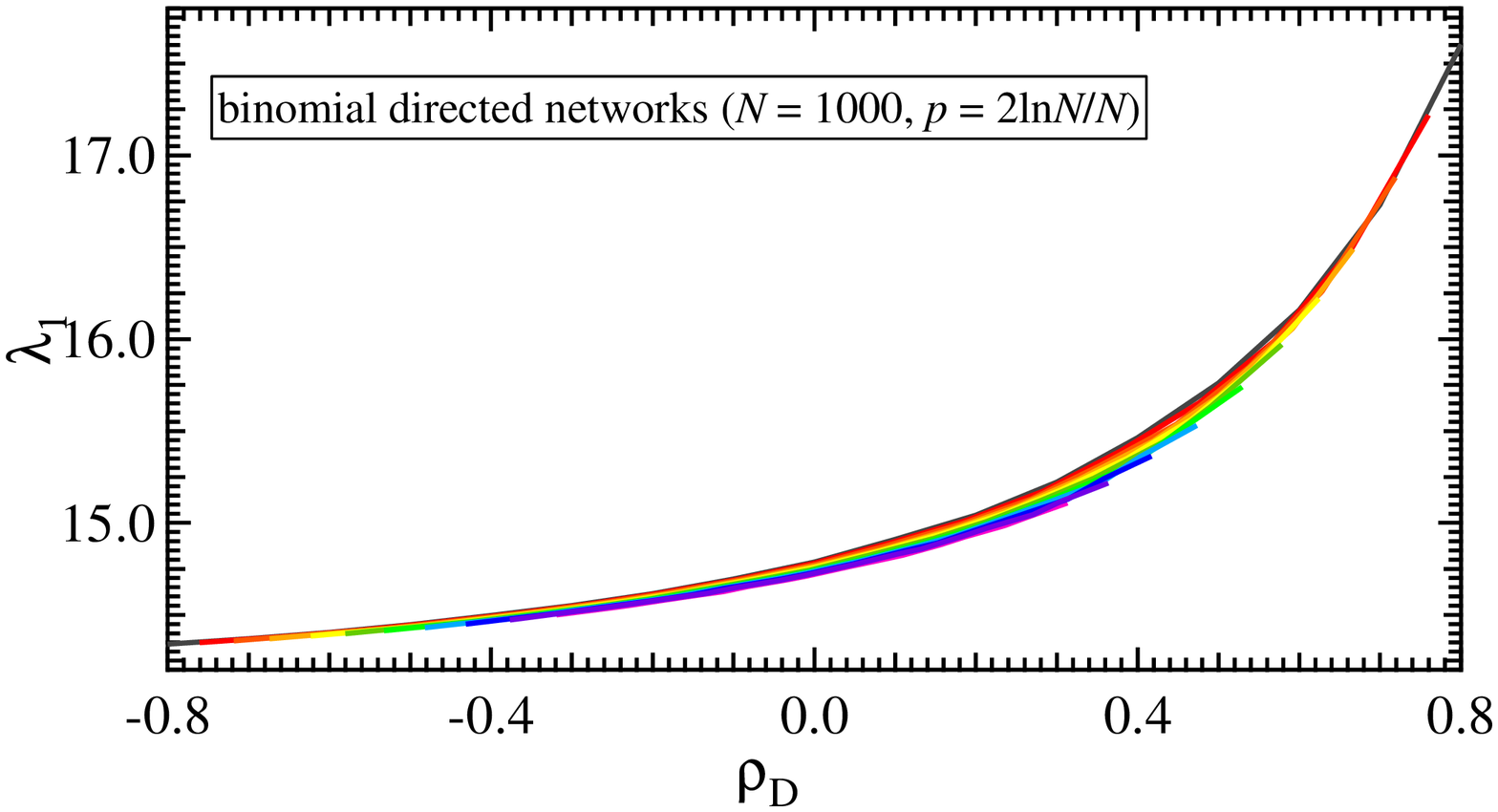}%
\\
(a) one network realization
\end{center}}}
%EndExpansion
&
%TCIMACRO{\FRAME{ihFU}{3.3425in}{1.74in}{-0.0104in}{\Qcb{(b) $10^2$ network
%realizaitons}}{}{exactass100times.eps}{\special{ language "Scientific Word";
%type "GRAPHIC";  display "USEDEF";  valid_file "F";  width 3.3425in;
%height 1.74in;  depth -0.0104in;  original-width 8.9015in;
%original-height 4.9882in;  cropleft "0";  croptop "1";  cropright "1";
%cropbottom "0";  filename 'exactass100times.eps';file-properties "XNPEU";}} }%
%BeginExpansion
\raisebox{0.0104in}{\parbox[b]{3.3425in}{\begin{center}
\includegraphics[
height=1.74in,
width=3.3425in
]%
{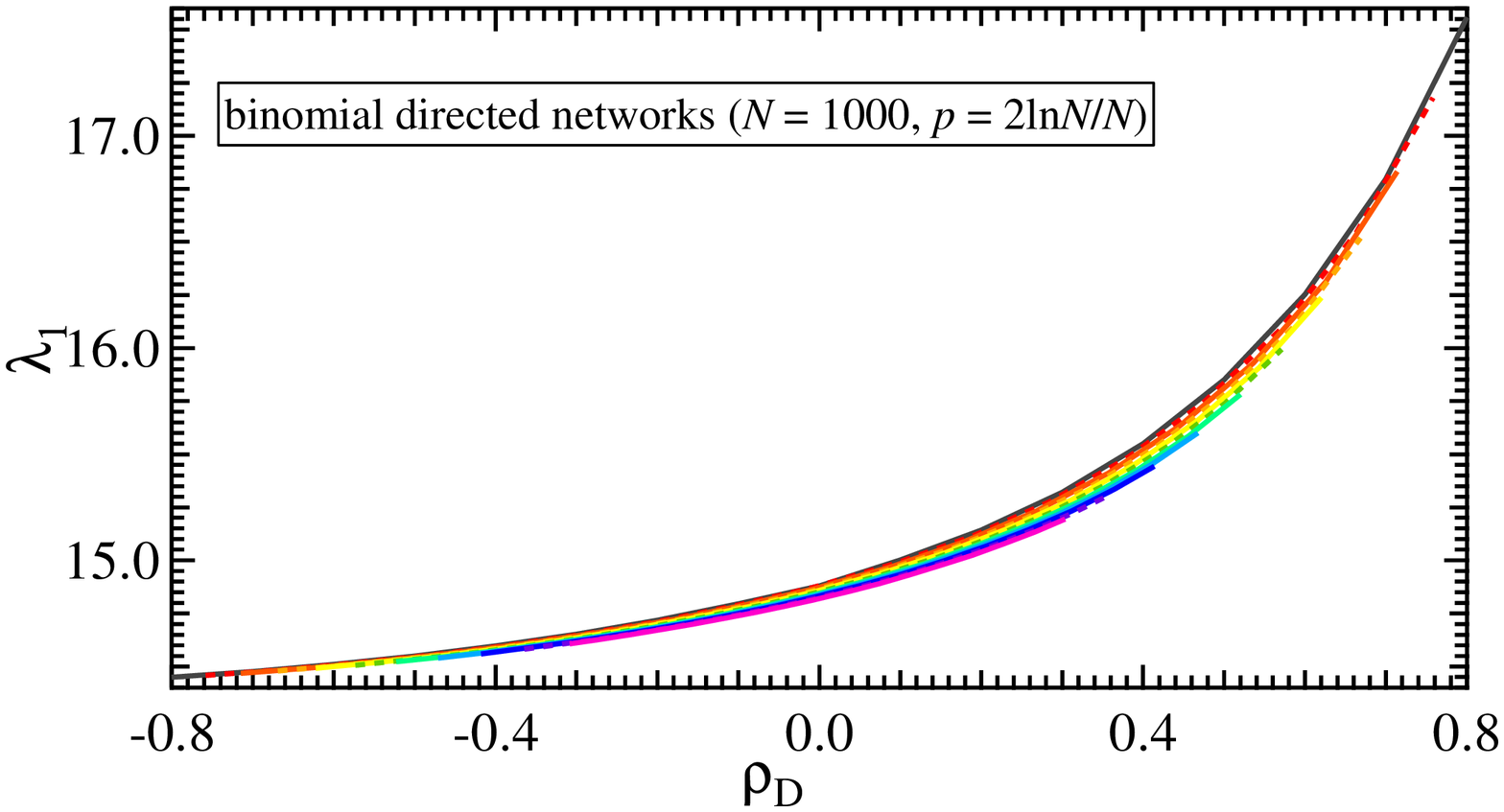}%
\\
(b) $10^2$ network realizaitons
\end{center}}}
%EndExpansion
\end{array}
$%
%TCIMACRO{\TeXButton{figure_7}{\caption
%{(Color online) Spectral radius as a function of the assortativity in directed binomial networks with $\xi
%$ from $0$ to $1$ with step $0.1$ is scatter plotted in different colorful (or grayscale) lines, from the gray (upper) line to the pink (lower) line.}%
%\label{Binomial_1}
%\end{figure}}}%
%BeginExpansion
\caption
{(Color online) Spectral radius as a function of the assortativity in directed binomial networks with $\xi
$ from $0$ to $1$ with step $0.1$ is scatter plotted in different colorful (or grayscale) lines, from the gray (upper) line to the pink (lower) line.}%
\label{Binomial_1}
\end{figure}%
%EndExpansion

\begin{figure}[h]%
\centering
\includegraphics[
height=1.9501in,
width=3.5319in
]%
{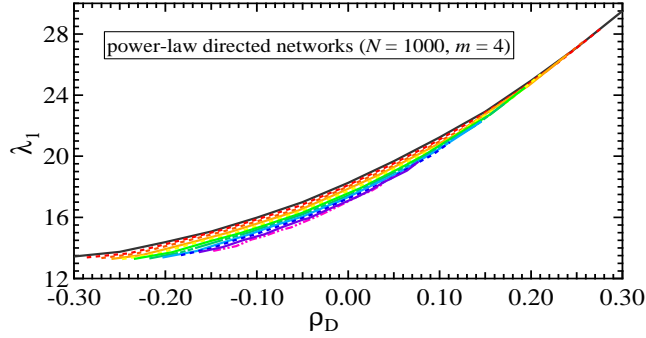}%
\caption{(Color online) Spectral radius as a function of the assortativity in
directed power-law networks with $\xi$ from $0$ to $1$ with step $0.1$,\ is
scatter plotted in different colorful (or grayscale) lines, from the gray
(upper) line to the pink (lower) line.}%
\label{spectraldecvsassBA}%
\end{figure}

$\bullet$\textbf{ Approach }$1$\textbf{: }First, we perform degree-preserving
rewiring on ER networks (or BA networks) to obtain a set of bidirectional
networks with assortativity $\rho_{D}$ from $-0.8$ to $0.8$ (or $-0.3$ to
$0.3$) with step $0.1$. Second, we alter the directionality\ $\xi$ of all
bidirectional networks with each assortativity using IOPRA. The directionality
$\xi$ is changed from $0$ to $1$ with step $0.1$. IOPRA randomizes network
connections, and thus pushes the assortativity of the resulting directed
network towards zero, if the original network has a non-zero assortativity.
Figure \ref{Binomial_1} plots the simulation results of one binomial network
realization and $10^{2}$ binomial network realizations. The simulation of one
realization is almost the same as the result of a large number of network
realizations, which points to \emph{almost sure behavior} \cite{Piet_almost
sure}. The results in the directed power-law networks are shown in Figure
\ref{spectraldecvsassBA}.%
%TCIMACRO{\FRAME{fhFU}{3.5319in}{1.9501in}{0pt}{\Qcb{(Color online) Spectral
%radius as a function of the assortativity in directed power-law networks with
%$\xi$ from $0$ to $1$ with step $0.1$,\ is scatter plotted in different
%colorful (or grayscale) lines, from the gray (upper) line to the pink (lower)
%line.}}{\Qlb{spectraldecvsassBA}}{effectofdirectionba.eps}%
%{\special{ language "Scientific Word";  type "GRAPHIC";  display "USEDEF";
%valid_file "F";  width 3.5319in;  height 1.9501in;  depth 0pt;
%original-width 8.0151in;  original-height 4.7824in;  cropleft "0";
%filename 'effectofdirectionBA.eps';file-properties "XNPEU";}} }%
%BeginExpansion
%EndExpansion

\begin{figure}
\centering
%EndExpansion
$%
\begin{array}
[c]{cc}%
%TCIMACRO{\FRAME{ihFU}{3.1695in}{1.7513in}{0in}{\Qcb{(a)}}{}%
%{binomial1000biggestdec1000.eps}{\special{ language "Scientific Word";
%type "GRAPHIC";  display "USEDEF";  valid_file "F";  width 3.1695in;
%height 1.7513in;  depth 0in;  original-width 8.8747in;
%original-height 4.9882in;  cropleft "0";  croptop "1";  cropright "1";
%filename 'Binomial1000biggestdec1000.eps';file-properties "XNPEU";}} }%
%BeginExpansion
{\parbox[b]{3.1695in}{\begin{center}
\includegraphics[
height=1.7513in,
width=3.1695in
]%
{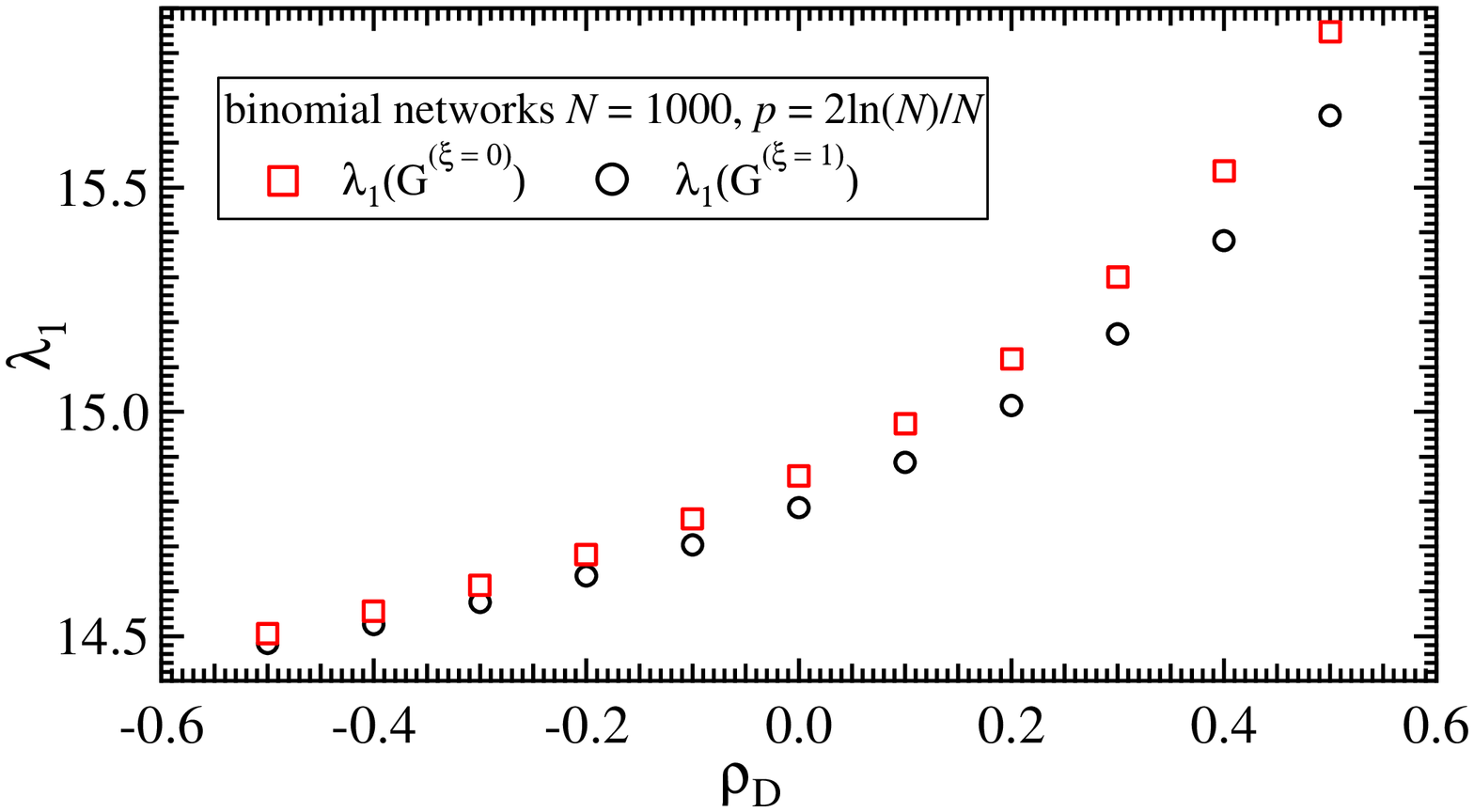}%
\\
(a)
\end{center}}}
%EndExpansion
&
%TCIMACRO{\FRAME{ihFU}{3.3217in}{1.7513in}{0in}{\Qcb{(b)}}{}%
%{powerlaw1000biggestdec1000.eps}{\special{ language "Scientific Word";
%type "GRAPHIC";  display "USEDEF";  valid_file "F";  width 3.3217in;
%height 1.7513in;  depth 0in;  original-width 8.9015in;
%original-height 4.9882in;  cropleft "0";  croptop "1";  cropright "1";
%filename 'powerlaw1000biggestdec1000.eps';file-properties "XNPEU";}} }%
%BeginExpansion
{\parbox[b]{3.3217in}{\begin{center}
\includegraphics[
height=1.7513in,
width=3.3217in
]%
{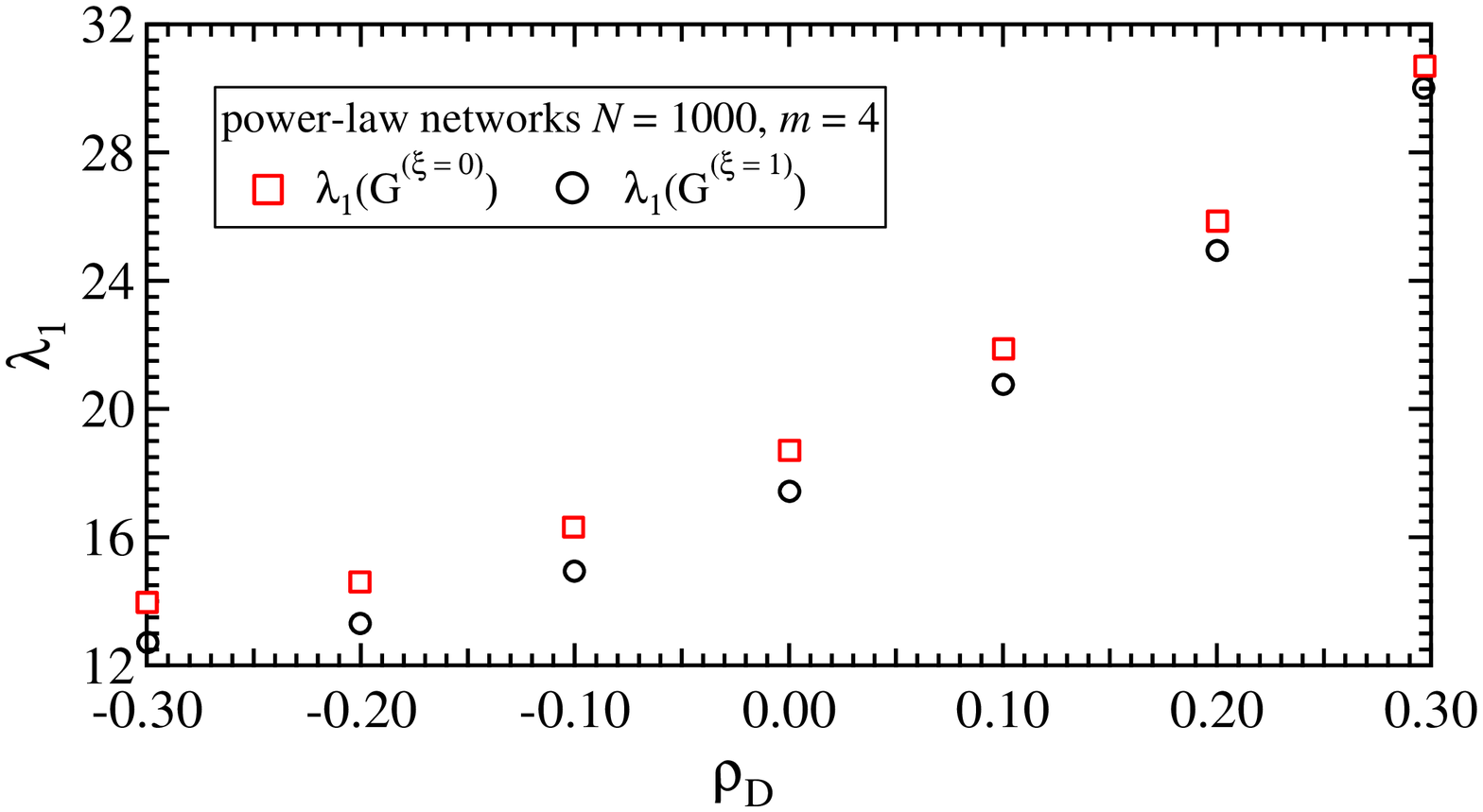}%
\\
(b)
\end{center}}}
%EndExpansion
\end{array}
$%
%TCIMACRO{\TeXButton{figure_8}{\caption
%{(Color online) Spectral radius as a function of the assoratativity (a) in directed binomial networks ($N=1000$, $p=2lnN/N$) and (b) in directed power-law networks ($N=1000$, $m=4$) for $10^{3}%
%$ network realizations.}\label{spectradecVSass}
%\end{figure}}}%
%BeginExpansion
\caption
{(Color online) Spectral radius as a function of the assoratativity (a) in directed binomial networks ($N=1000$, $p=2lnN/N$) and (b) in directed power-law networks ($N=1000$, $m=4$) for $10^{3}%
$ network realizations.}\label{spectradecVSass}
\end{figure}

$\bullet$\textbf{ Approach }$2$\textbf{: }First, we generate ER networks (or
BA networks) $G^{(\xi=0)}$ whose directionality $\xi=0$. Second, we apply
IOPRA to ER (or BA) networks $G^{(\xi=0)}$ to generate directed binomial
networks (or directed power-law networks) $G^{(\xi=1)}$ with directionality
$\xi=1$. Then, we change the assortativity of ER networks (or BA networks)
$G^{(\xi=0)}$\ and directed binomial networks (or directed power-law networks)
$G^{(\xi=1)}$\ by degree preserving rewiring and in- and out- degree
preserving rewiring, respectively, without changing the directionality. Note
that the in- and out- degree preserving rewiring can be applied not only to
change the directionality, but also to change the assortativity. Figure
\ref{spectradecVSass} plots the spectral radius of the networks $G^{(\xi=0)}$
and $G^{(\xi=1)}$ as a function of the assortativity.%

%TCIMACRO{\TeXButton{figure_8}{\begin{figure}
%\centering}}%
%BeginExpansion
%EndExpansion

The two approaches both change the assortativity and the directionality of the
networks, while, the order of change is different: Approach $1$\ changes the
assortativity firstly and then the directionality; Approach 2 is the opposite.
Figures \ref{Binomial_1}, \ref{spectraldecvsassBA} and \ref{spectradecVSass},
show that the spectral radius $\lambda_{1}$ always decreases with the
directionality $\xi$ when the networks have the same degree distribution and
the same assortativity $\rho_{D}$. Moreover, the degree distribution of the
network also influences the change range of the spectral radius $\lambda_{1}%
$\ with $\xi$. The decrement of the spectral radius $\lambda_{1}$ with $\xi$
increases with the assortativity in directed binomial networks (see Figures
\ref{Binomial_1} and \ref{spectradecVSass} (a)). On the contrary, the
decrement of the spectral radius $\lambda_{1}$ with $\xi$ goes down with the
assortativity in directed power-law networks (see Figures
\ref{spectraldecvsassBA} and \ref{spectradecVSass} (b)). Furthermore, the
decrease of the spectral radius in directed power-law networks is larger than
that in directed binomial networks, when the assortativity is zero. Many
real-world networks are directed power-law networks, where $\lambda_{1}%
$\ could possibly be tuned within a large range by controlling the
directionality in real-world networks.

Summarizing, the spectral radius $\lambda_{1}$ decreases with the
directionality $\xi$ when the assortativity remains constant. In order to
protect the network from virus spreading via increasing the epidemic
threshold, while maintaining the degree distribution and the assortativity,
increasing the directionality of networks is recommended. Meanwhile, the
spectral gap and the algebraic connectivity are also increased, which means
that the topological robustness is also enhanced in return.

\section{Conclusions}

In this work, two algorithms to generate directed networks with a given
directionality $\xi$ are proposed. This allows us to study the influence of
the directionality $\xi$ on the spectral properties of networks. The spectral
radius $\lambda_{1}$, which is the inverse of the SIS\ NIMFA epidemic
threshold $\tau_{c}^{(1)}$, is studied in directed networks. A universal
observation is that, the spectral radius decreases with the directionality
when the degree distribution and the assortativity of the network is
preserved. We may, thus, increase the epidemic threshold to suppress the virus
spread via increasing the directionality of the network. The possible range to
increase the epidemic threshold is relatively large in\ directed binomial
networks with a high assortativity and directed power-law networks with a low
assortativity. The variance of the components of the principal eigenvector
decreases with the directionality, which indicates that the influence of each
node on the spectral radius is similar in networks with a high directionality.
Moreover, the spectral gap and the algebraic connectivity increase with the
directionality, implying that an increase of the directionality enhances the
connectivity of the network. Furthermore, we observe that the spectral gap
increases faster with the directionality in IOPRA than in LRA directed
binomial networks, on the contrary, the algebraic connectivity increases with
the directionality faster in LRA than in IOPRA directed binomial networks.
This observation may be due to the fact that the in- and out- degree of each
node could be different in LRA directed binomial networks, while, are exactly
the same in IOPRA directed binomial networks. The influence of the difference
between in- and out- degree of nodes on spectral properties for directed
power-law networks remains an open question.

\textbf{Acknowledgements}

This work has been partially supported by the European Commission within the
framework of the CONGAS project FP7-ICT-2011-8-317672 and the China
Scholarship Council (CSC).

\ \ \ \ \ \ \ \ \ \ \ \ \ \ \ \ \ \ \ \ \ \ \ \ \ \ \ \ \ \ \ \ \ \ \ \ \ \ \ \ \ \ \ \ \ \ \ \ \ \ \ \ \ \ \ \ \ \ \ \ \ \ \ \ \ {\LARGE Appendix\bigskip
}

{\Large A. Introduction of real-world networks\medskip}

{\large 1. Enron:}

This data set was made public by the Federal Energy Regulatory Commission
during its investigations: it is a partially anonymised corpus of e-mail
messages exchanged by some Enron employees (mostly part of the senior
management). This data set is a directed graph, whose nodes represent people
and with an arc from x to y whenever y was the recipient of (at least) a
message sent by x.\smallskip

{\large 2. Ljournal-2008:}

LiveJournal is a virtual-community social site started in 1999: nodes are
users and there is an arc from x to y if x registered y among his friends. It
is not necessary to ask y permission, so the graph is directed. This graph is
the snapshot used by Chierichetti, Flavio, et al. in "On compressing social
networks." Proceedings of the 15th ACM SIGKDD international conference on
Knowledge discovery and data mining. ACM, 2009, and was kindly provided by the
authors.\smallskip

{\large 3. Twitter-2010}

Twitter is a website, owned and operated by Twitter Inc., which offers a
social networking and microblogging service, enabling its users to send and
read messages called tweets. Tweets are text-based posts of up to 140
characters displayed on the user's profile page. This is a crawl presented by
Kwak, Haewoon, et al. in "What is Twitter, a social network or a news media?",
Proceedings of the 19th international conference on World wide web. ACM, 2010.
Nodes are users and there is an arc from x to y if y is a follower of x. In
other words, arcs follow the direction of tweet transmission.\smallskip

{\large 4. Word Association-2011}

The Free Word Association Norms Network is a directed graph describing the
results of an experiment of free word association performed by more than 6000
participants in the United States: its nodes correspond to words and arcs
represent a cue-target pair (the arc x-%
%TCIMACRO{\TEXTsymbol{>}}%
%BeginExpansion
$>$%
%EndExpansion
y means that the word y was output by some of the participants based on the
stimulus x).\smallskip

{\large 5. WWW networks}

The networks, "cnr-2000", "in-2004", "eu-2005", "uk-2007-05@100000" and
"uk-2007-05@1000000" are small WWW networks that were crawled from the
Internet. The "cnr-2000" is crawled from the Italian CNR domain. A small crawl
of the .in domain performed for the Nagaoka University of Technology is in
data "in-2004". The "eu-2005" is a small crawl of the .eu domain. This network
"uk-2007-05@100000" and "uk-2007-05@1000000" have been artificially generated
by combining twelve monthly snapshot of the .uk domain and collected for the
DELIS project.\smallskip

{\Large B. Eigenvalues of the directed networks\medskip}

\begin{figure}[h]%
\centering
\includegraphics[
height=2.2857in,
width=4.2462in
]%
{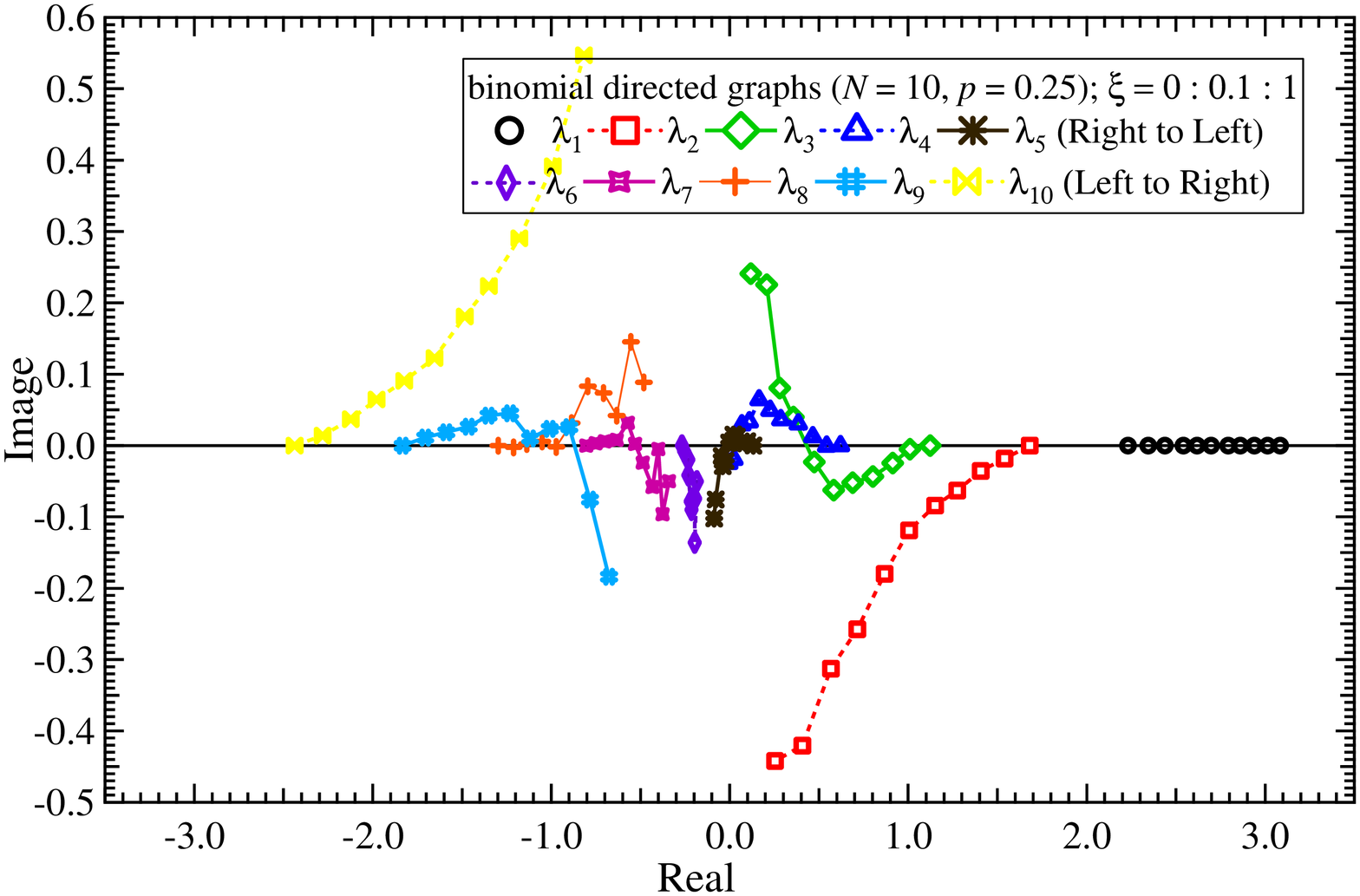}%
\caption{Change of the eigenvalues with the directionality. When the
directionality increases, the real parts of the eigenvalues tend to $0$.}%
\label{Eigenvalues}%
\end{figure}

The spectral radius\ $\lambda_{1}$\ and the spectral gap $(\lambda_{1}%
-\lambda_{2})$ are considered as important metrics for the percolation
processes on networks. Here we also present all eigenvalues in directed
networks in a Image-Real figure. The eigenvalues are calculated on $10^{3}$
simulation realizations. The changes of the eigenvalues $\lambda_{i}$ with the
directionality $\xi$ from $0$ to $1$ with step $0.1$ in directed binomial
networks ($N=10$, $p=0.25$) are shown on Figure \ref{Eigenvalues}.
Surprisingly, the real part of all eigenvalues tends to $0$, when the
directionality $\xi$ increases.%

%TCIMACRO{\FRAME{fhFU}{4.2462in}{2.2857in}{0pt}{\Qcb{Change of the eigenvalues
%with the directionality. When the directionality increases, the real parts of
%the eigenvalues tend to $0$.}}{\Qlb{Eigenvalues}}{eigenvalues.eps}%
%{\special{ language "Scientific Word";  type "GRAPHIC";  display "USEDEF";
%valid_file "F";  width 4.2462in;  height 2.2857in;  depth 0pt;
%original-width 11.291in;  original-height 7.6268in;  cropleft "0";
%filename 'eigenvalues.eps';file-properties "XNPEU";}} }%
%BeginExpansion
%EndExpansion
\bigskip

{\Large C. Algorithms (see Algorithm \ref{IOPRAAlgo}{\LARGE \ }and Algorithm
\ref{LRAAlgo})}

\floatstyle{ruled} \newfloat{algorithm}{htbp}{loa}
\floatname{algorithm}{Algorithm}
%\floatname{algorithm}{Procedure}%
%TCIMACRO{\TeXButton{Algorithm}{\begin{algorithm}}}%
%BeginExpansion
\begin{algorithm}%
%EndExpansion

\caption{$IOPRA(G,\ \xi)$} \label{IOPRAAlgo}

\begin{algorithmic}[1]
\STATE $Create\ a\ bidirectional\ network\ G(N,\ L);$
\STATE $Save\ network\ G(N,L)\ as\ G _{s}\ and\ calculate\ the\ directionality\ \xi _{s}\ of\ network\ G _{s}; $
\WHILE{$\left\vert \xi _{s}-\xi \right\vert >10^{-5}$}
\STATE $Randomly\ select\ two\ unidirectional\ links\ i\rightarrow j\ and\ k\rightarrow l\ associated\ with\ the\ four\ nodes\ i,\ j,\ k,\ l;$
\STATE $Rewire\ the\ link\ pair\ i\rightarrow j\ and\ k\rightarrow l\ into\ i\rightarrow l\ and\ k\rightarrow j.\ The\ new\ network\ G _{n}\ is\ obtained;$
\STATE $calculate\ the\ directionality\ \xi _{n}\ of\ the\ network\ G _{n}; $
\IF{$\left\vert \xi _{s}-\xi \right\vert >\left\vert \xi _{n}-\xi \right\vert $}
\STATE $G _{s}\leftarrow G _{n};$
\STATE $\xi _{s}\leftarrow \xi _{n};$
\ELSE
\STATE $give\ up\ this\ rewired\ node\ pair;$
\ENDIF
\ENDWHILE
\RETURN $G _{s}$
\end{algorithmic}
%

%TCIMACRO{\TeXButton{Algorithm}{\end{algorithm}}}%
%BeginExpansion
\end{algorithm}%
%EndExpansion

\floatstyle{ruled} \newfloat{algorithm}{htbp}{loa}
\floatname{algorithm}{Algorithm}
%\floatname{algorithm}{Procedure}%
%TCIMACRO{\TeXButton{Algorithm}{\begin{algorithm}}}%
%BeginExpansion
\begin{algorithm}%
%EndExpansion

\caption{$LRA(G,\ \xi)$} \label{LRAAlgo}

\begin{algorithmic}[1]
\STATE $Create\ a\ bidirectional\ network\ G(N,\ L);$
\STATE $Randomly\ choose\ \xi\ percentage\ of\ bidirectional\ link\ pairs;$
\STATE $Randomly\ choose\ one\ unidirectional\ link\ from\ each\ link\ pair;$
\STATE $Randomly\ reset\ the\ chosen\ unidirectional\ links\ to\ the\ locations\ without\ any\ link;$
\STATE $Save\ the\ new\ network\ as\ G _{s};$
\RETURN $G _{s}$
\end{algorithmic}
%

%TCIMACRO{\TeXButton{Algorithm}{\end{algorithm}}}%
%BeginExpansion
\end{algorithm}%
%EndExpansion

\begin{figure}[b]
\centering
%EndExpansion
$%
\begin{array}
[c]{cc}%
%TCIMACRO{\FRAME{itbpF}{3.2595in}{2.1335in}{0in}{}{}{g10000small.eps}%
%{\special{ language "Scientific Word";  type "GRAPHIC";  display "USEDEF";
%valid_file "F";  width 3.2595in;  height 2.1335in;  depth 0in;
%original-width 9.0676in;  original-height 6.0727in;  cropleft "0";
%filename 'G10000small.eps';file-properties "XNPEU";}} }%
%BeginExpansion
{\includegraphics[
height=2.1335in,
width=3.2595in
]%
{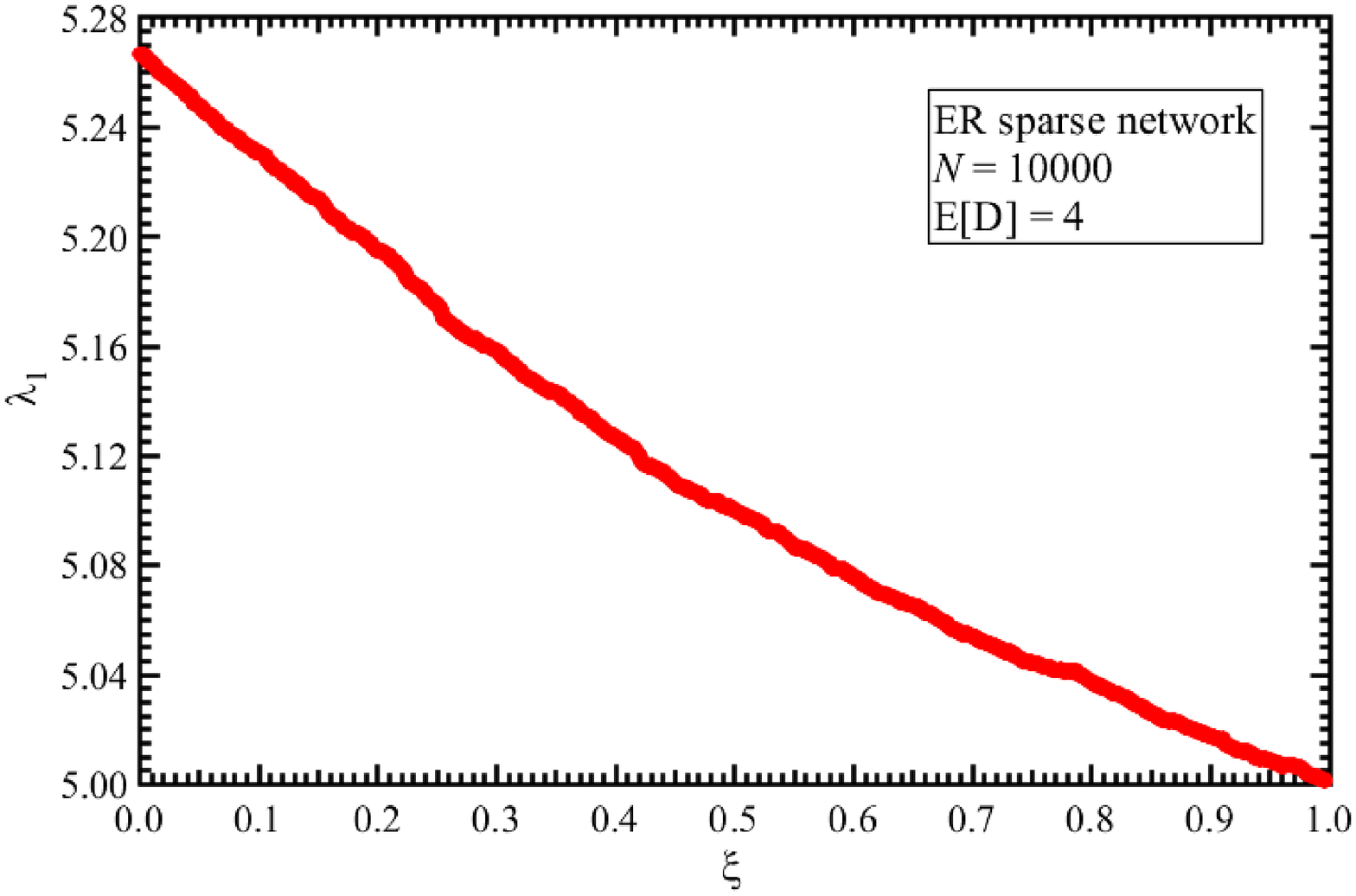}%
}
%EndExpansion
&
%TCIMACRO{\FRAME{itbpF}{3.2292in}{2.1162in}{0in}{}{}{g100000small.eps}%
%{\special{ language "Scientific Word";  type "GRAPHIC";  display "USEDEF";
%valid_file "F";  width 3.2292in;  height 2.1162in;  depth 0in;
%original-width 9.0278in;  original-height 6.045in;  cropleft "0";
%filename 'G100000small.eps';file-properties "XNPEU";}} }%
%BeginExpansion
{\includegraphics[
height=2.1162in,
width=3.2292in
]%
{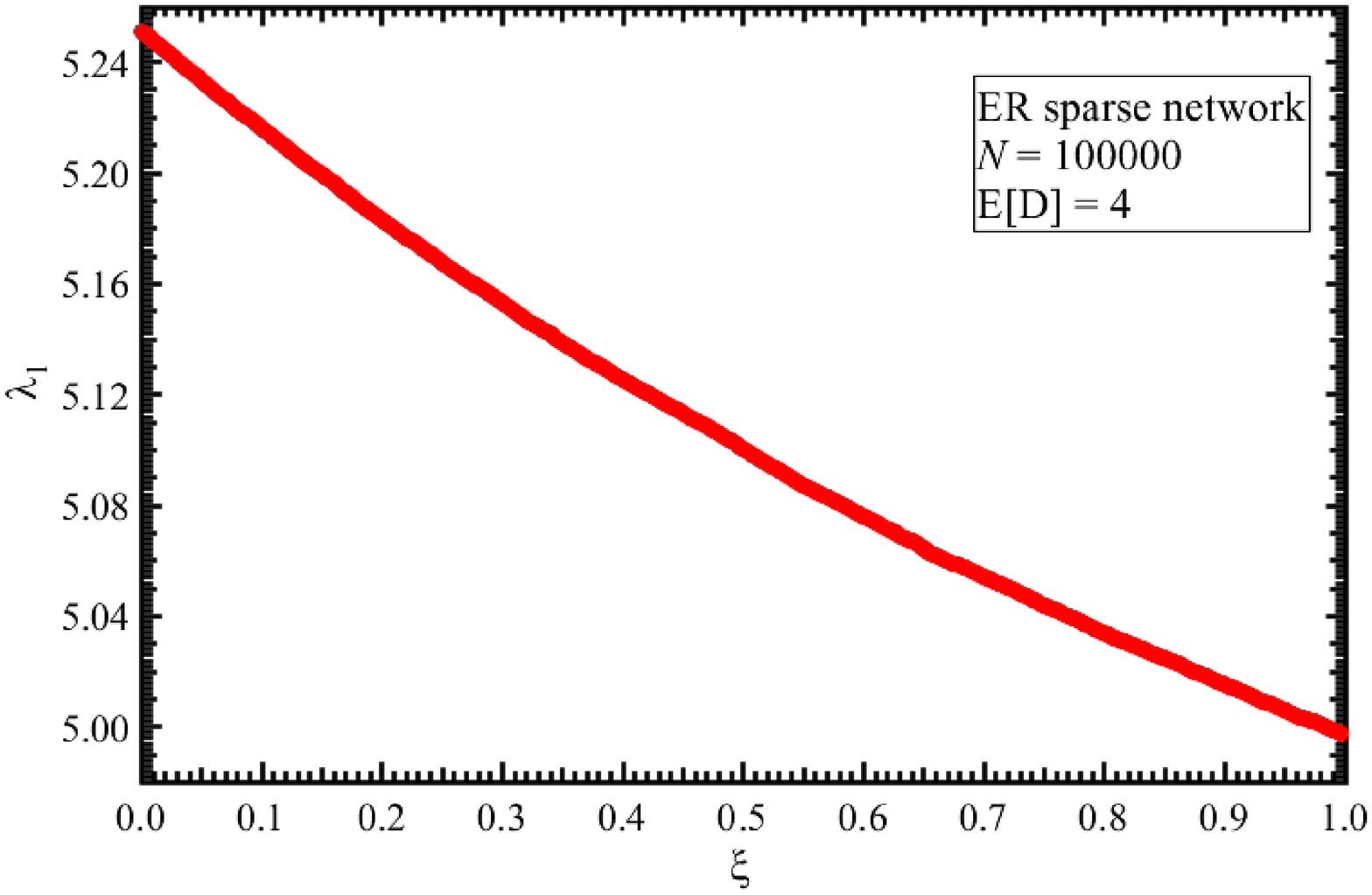}%
}
%EndExpansion
\end{array}
$%
%TCIMACRO{\TeXButton{figure_8}{\caption
%{(Color online) Spectral radius as a function of the directionality in large, sparse directed binomial networks.}%
%\label{large_sparse}
%\end{figure}}}%
%BeginExpansion
\caption
{(Color online) Spectral radius as a function of the directionality in large, sparse directed binomial networks.}%
\label{large_sparse}
\end{figure}

{\Large D. Spectral radius of large sparse networks (see Figure
\ref{large_sparse})}%

%TCIMACRO{\TeXButton{figure_8}{\begin{figure}[b]
%\centering}}%
%BeginExpansion
%EndExpansion

\end{document}